\documentclass[sigconf, screen]{acmart}
\pdfoutput=1
\AtBeginDocument{%
  \providecommand\BibTeX{{%
    \normalfont B\kern-0.5em{\scshape i\kern-0.25em b}\kern-0.8em\TeX}}}

\setcopyright{acmcopyright}
\copyrightyear{2024}
\acmYear{2024}
\acmDOI{XXXXXXX.XXXXXXX}

\acmConference[CHI'24]{ACM CHI conference on Human Factors in Computing Systems}{May 11-16,
  2024}{Honolulu, USA}
\acmBooktitle{CHI'24: ACM CHI conference on Human Factors in Computing Systems,
 April, 2024} 
\acmPrice{15.00}
\acmISBN{978-1-4503-XXXX-X/18/06}

\usepackage{booktabs}
\usepackage{multirow}
\usepackage{colortbl}
\usepackage{xspace}
\usepackage{graphicx}
\usepackage{xcolor}
\usepackage[most]{tcolorbox}
\usepackage{subcaption}
\usepackage{soul}
\usepackage{fontawesome}
\usepackage{tabularx}
\usepackage{changepage}
\usepackage{array}
\usepackage[normalem]{ulem}
\useunder{\uline}{\ul}{}
\usepackage{hyperref}
\usepackage{ulem}

\newif\ifcomment

\ifcomment
  \newcommand{\gao}[1]{\textcolor[rgb]{0.6,0,0.2}{GJ: #1}}
  \newcommand{\simon}[1]{\textcolor[rgb]{0,0.8,0.4}{Simon: #1}}
  \newcommand{\added}[1]{\textcolor[rgb]{0.2,0.5,0.8}{#1}}
  \newcommand{\deleted}[1]{\sout{\textcolor[rgb]{0.8,0.8,0.8}{#1}}}

\else
  \newcommand{\gao}[1]{}
  \newcommand{\simon}[1]{}
  \newcommand{\added}[1]{\textcolor{black}{#1}}
  \newcommand{\deleted}[1]{}
\fi

\newcommand{\diffcoder}{{CollabCoder}\xspace}

\newcommand{\affiliationline}[2]{\institution{#1, #2}\country{}}
\newcolumntype{L}{>{\raggedright\arraybackslash}X}

\definecolor{figma_green}{HTML}{72c87a}
\definecolor{figma_blue}{HTML}{8fb5f9}
\definecolor{figma_orange}{HTML}{f18f6d}
\definecolor{figma_yellow}{HTML}{f5c242}
\definecolor{light_grey}{rgb}{0.9, 0.9, 0.9}
\definecolor{figma_purple}{HTML}{000000}
\sethlcolor{light_grey}

\newcommand*\circled[1]{\tikz[baseline=(char.base)]{\node[shape=circle,draw,inner sep=0.5pt] (char) {#1};}}

\newcommand{\numberedphraselowcase}[2]{\circled{#1} #2}

\begin{document}

\title[\diffcoder]{\diffcoder: A Lower-barrier, Rigorous Workflow for \added{Inductive} Collaborative Qualitative Analysis with Large Language Models}

\author{Jie Gao}
\affiliation{%
  \affiliationline{Singapore University of Technology and Design}{Singapore}
}
\affiliation{%
    \affiliationline{Singapore-MIT Alliance for Research and Technology}{Singapore}
}
\email{gaojie056@gmail.com}

\author{Yuchen Guo}
\affiliation{
   \affiliationline{Singapore University of Technology and Design}{Singapore}}
\email{yuchen_guo@mymail.sutd.edu.sg}

\author{Gionnieve Lim}
\affiliation{
   \affiliationline{Singapore University of Technology and Design}{Singapore}}
\email{gionnieve_lim@mymail.sutd.edu.sg}

\author{Tianqin Zhang}
\affiliation{
   \affiliationline{Singapore University of Technology and Design}{Singapore}}
\email{tianqin_zhang@mymail.sutd.edu.sg}

\author{Zheng Zhang}
\affiliation{
   \affiliationline{University of Notre Dame}{USA}}
\email{zzhang37@nd.edu}

\author{Toby Jia-Jun Li}
\affiliation{
   \affiliationline{University of Notre Dame}{USA}}
\email{toby.j.li@nd.edu}

\author{Simon Tangi Perrault}

\affiliation{
   \affiliationline{Singapore University of Technology and Design}{Singapore}}
\email{simon_perrault@sutd.edu.sg}

\renewcommand{\shortauthors}{Jie Gao, Yuchen Guo, Gionnieve Lim, Tianqin Zhang, Zheng Zhang, Toby Jia-Jun Li, Simon Tangi Perrault}

\begin{abstract}
Collaborative Qualitative Analysis (CQA) can enhance qualitative analysis rigor and depth by incorporating varied viewpoints. Nevertheless, ensuring a rigorous CQA procedure itself can be both demanding and costly.
To lower this bar, we \added{take a theoretical perspective to design the \diffcoder workflow}, that integrates Large Language Models (LLMs) into key \textit{inductive} CQA stages: independent open coding, iterative discussions, and final codebook creation. In the open coding phase, \diffcoder offers AI-generated code suggestions and records decision-making data. During discussions, it promotes mutual understanding by sharing this data within the coding team and using quantitative metrics to identify coding (dis)agreements, aiding in consensus-building. In the code grouping stage, \diffcoder provides primary code group suggestions, lightening the cognitive load of finalizing the codebook. \added{A 16-user evaluation confirmed the effectiveness of \diffcoder, demonstrating its advantages over existing software and providing empirical insights into the role of LLMs in the CQA practice.}
\end{abstract}



\begin{teaserfigure}
  \includegraphics[width=\textwidth]{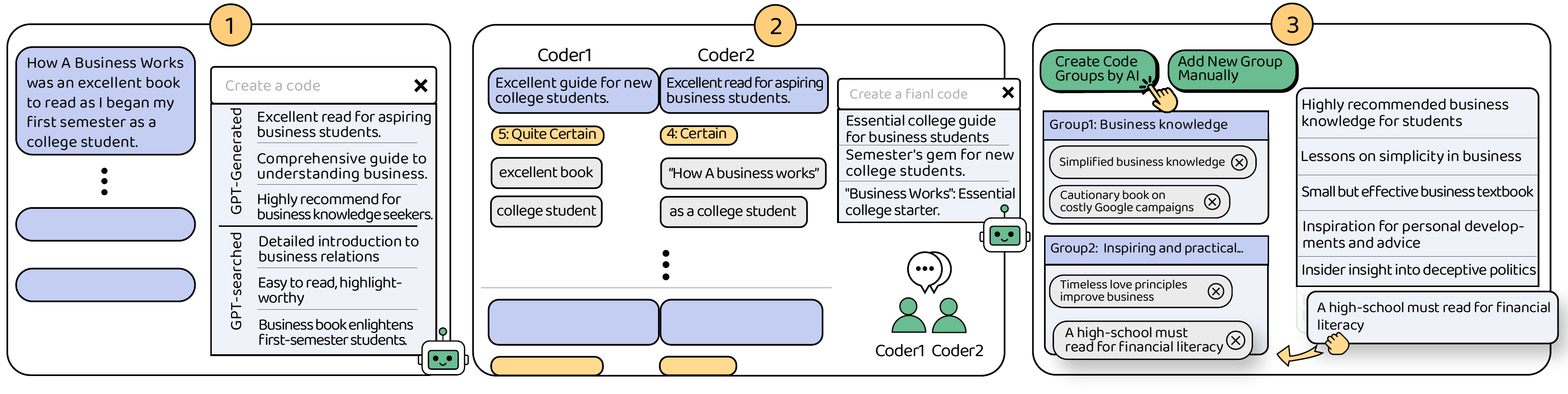}
  \caption{\diffcoder, An Integrated Workflow for Collaborative Qualitative Analysis. The workflow consists of three key stages: 1) Independent Open Coding, facilitated by on-demand code suggestions from LLMs, yielding initial codes; 2) Iterative Discussion, focusing on conflict mediation within the coding team, producing a list of agreed-upon code decisions; 3) Codebook Development, where code groups may be formed through LLM-generated suggestions, based on the list of decided codes.}
  \Description{Main Interface of \diffcoder.}
  \label{fig:teaser}
\end{teaserfigure}

\maketitle

\section{Introduction}

\begin{figure}[!t]
    \centering
    \includegraphics[width=0.42\textwidth]{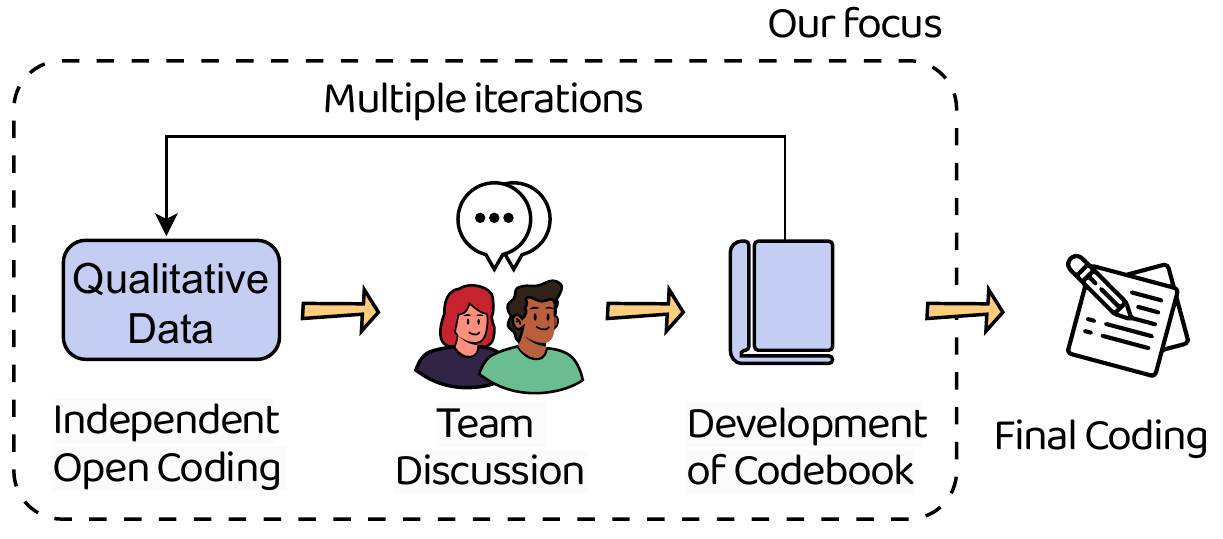}
    \caption{Collaborative Qualitative Analysis (CQA)~\cite{corbin2009basic, corbin1990grounded, richards2018practical} is an iterative process involving multiple rounds of iteration among coders to reach a final consensus. Our goal with \diffcoder is to assist users across key stages of the CQA process.}
    \label{fig:intro:iteration}
\end{figure}

Rigor and in-depth interpretation are primary objectives in qualitative analysis \cite{watkins2017rapid, maher2018ensuring}. Collaborative Qualitative Analysis (CQA) underscores this by mandating researchers to code individually and converge on interpretations through iterative discussions~\cite{cornish2013collaborative, anderson2016all, hall2005qualitative, richards2018practical, mcdonald2019reliability} (see Figure \ref{fig:intro:iteration}). Such a method is instrumental in preserving rigor~\cite{richards2018practical} and facilitating a richer, more nuanced grasp of data interpretation~\cite{anderson2016all}. 

However, adhering strictly to the CQA's prescribed workflow, which is integral for achieving both rigor and depth goals, poses challenges due to its \textit{associated time and labor costs} as well as \textit{inherent complexity}.
\added{For the former issue, a primary reason is that the iterative nature of CQA requires the involvement and coordination of many coders~\cite{ganji2018ease, gao2023coaicoder}, but the conventional CQA tools such as MaxQDA, NVivo, and Google Docs/Sheets are not specifically designed for this aspect. They necessitate additional team coordination steps~\cite{malone1994coordination, entin1999adaptive}, like document downloading, data sharing in the team, data importing, manual searching, and crafting codebook tables. For the latter issue, the complexity of the CQA process, which involves multiple steps with specific requirements for each, presents a considerable entry barrier for those less experienced or unfamiliar with CQA standards like graduate students, early-career researchers, and diverse research teams~\cite{cornish2013collaborative, richards2018practical}. For instance, Atlas.ti Web lacks an independent coding space. This absence means that the coding process is always visible and can potentially influence others' open coding~\cite{gao2023coaicoder}, which might lead to confusion or incorrect practices. Despite this challenge, most software is specifically engineered to support basic functions, such as proposing codes. They typically lack a comprehensive and holistic theoretical framework that could provide more effective assistance.
This limitation leads to much confusion, even among those well-versed in CQA theories, compelling them to opt for independent coding methods in exchange for efficiency, resulting in fewer interactive discussions, diminished coding rigor and depth, and ultimately, the risk of the outcomes reflecting the individual coder’s inherent biases~\cite{cornish2013collaborative, anderson2016all}.}

\added{Current HCI researchers are mainly focusing on addressing effort-intensive challenges and have developed specialized tools to streamline various aspects of the CQA process.}
For example, Zade et al.~\cite{zade2018conceptualizing} suggested enabling coders to order different states of disagreements by conceptualizing disagreements in terms of tree-based ranking metrics of diversity and divergence. Aeonium~\cite{drouhard2017aeonium} allows coders to highlight ambiguity and inconsistency and offers features to navigate through and resolve them. 
With the growing prevalence of AI, Gao et al.~\cite{gao2023coaicoder} underscores the potential of AI in CQA through CoAIcoder, suggesting that AI agents that provide code suggestions based on teams' coding histories accelerate collaborative efficiency and foster a shared understanding at the early stage of coding.

\added{With the advancements of Large Language Models (LLMs)\footnote{In this paper, AI and LLMs are used interchangeably to refer to the broader field of Artificial Intelligence, specifically large language models. GPT, as an example of a large language model, specifically refers to products developed by OpenAI, such as ChatGPT.} like GPT-3.5 and GPT-4\footnote{\url{https://openai.com/blog/introducing-chatgpt-and-whisper-apis}}, they have been pivotal in enhancing qualitative analysis due to the exceptional abilities in understanding and generating text. Atlas.ti Web, a commercial platform for qualitative analysis, integrated OpenAI's GPT model on March 28, 2023\footnote{\url{https://atlasti.com/ai-coding-powered-by-openai}}. This integration offers functionalities like one-click code generation and AI-driven code suggestions, significantly streamlining the coding process. LLMs also assist in the deductive coding of large-scale datasets~\cite{xiao2023supporting} and are being explored for their potential to replace human coders in discerning subtle data nuances~\cite{byun2023dispensing}, among other applications. This integration serves as a key factor in reducing the labor intensity typically associated with traditional qualitative analysis.
}

While this existing research provides valuable insights into various facets of CQA, there has been little emphasis on creating a streamlined workflow to bolster the rigorous CQA process. 
\added{Building upon well-accepted CQA steps that are deeply rooted in Grounded Theory ~\cite{clark1991grounding} and Thematic Analysis ~\cite{maguire2017doing},} we aim to address this gap by presenting a holistic, one-stop solution that enhances the CQA process, with an emphasis on the \textit{inductive} qualitative analysis, central to the development of the codebook and coding schema. 
\textbf{Our primary objective is to lower the bar of adherence to the rigorous CQA process}, \added{thereby providing a potential for enhancing the quality of qualitative interpretation~\cite{collins2018international} with controllable and manageable effort.}

To this end, we introduce \diffcoder, a CQA workflow that integrates LLMs for the development of code schemes.
Primarily, \diffcoder features interfaces tailored to a three-stage CQA workflow, aligned with standard CQA protocols. It facilitates real-time data synchronization and centralized management, obviating the need for intricate data exchanges among coders. The platform offers both individual and shared workspaces, facilitating seamless transitions between personal and collaborative settings at various stages. The shared workspace contains collective decision-making data and quantitative metrics, essential for addressing code discrepancies. Beyond basic functionalities, \diffcoder integrates GPT to achieve multiple goals within the workflow: 1) providing automated code suggestions to streamline open codes development; 2) aiding the conversion of open codes into final code decisions; and 3) \added{providing initial versions of code groups, derived from these code decisions, for coders to further refine and adjust.}

\added{We conducted an evaluation of the \diffcoder system, addressing the following key questions: }

\begin{itemize}
    \item \added{\textbf{RQ1.} Can \diffcoder's workflow support qualitative coders to conduct CQA effectively?}
    \item \added{\textbf{RQ2.} How does \diffcoder compare to currently available tools like Atlas.ti Web?}
    \item \added{\textbf{RQ3.} How can the design of the \diffcoder workflow be improved?}
\end{itemize}

Our evaluation of \diffcoder demonstrated its user-friendliness, particularly for beginners navigating the CQA workflow (75\%+ participants agree or strongly agree on \textit{"easy to use"} and \textit{"learn to use quickly"}). It effectively supports coding independence, fosters an understanding of (dis)agreements, and helps in building a shared understanding within teams (75\%+ participants agree or strongly agree that \diffcoder helps to \textit{"identify (dis)agreements"}, \textit{"understand others' thoughts"}, \textit{"resolve disagreements"} and \textit{"understand the current level of agreement"}). Additionally, \diffcoder optimizes the discussion phase by allowing code pairs to be resolved in a single dialogue, in contrast to Atlas.ti Web where only a few codes are typically discussed. This minimizes the need for multiple discussion rounds, thereby boosting collaborative efficiency.
Regarding the role GPT plays in the CQA workflow, we emphasize the need to balance LLM capabilities with user autonomy, especially when GPT acts as a \textit{"suggestion provider"} during the initial phase. In the discussion stages, GPT functions as a \textit{"mediator"} and \textit{"facilitator"}, aiding in efficient and equitable team decision-making as well as in the formation of code groups. Finally, we discuss the limitations of our system based on certain assumptions (e.g., two-person coding teams, fixed data units) and suggest potential improvements to address these constraints.

Our study consequently paves the way for LLMs-powered (C)QA tools and also uncovers critical challenges and insights for both human-AI and human-human interactions within the context of qualitative analysis. We make the following contributions:

\begin{enumerate}
\item Outlining design guidelines, informed by both theories and practices, that shaped the development of \diffcoder and may inspire future AI-assisted CQA system designs.
\item Developing \diffcoder that incorporates LLMs into different steps of the inductive CQA workflow, enabling coders to seamlessly transition from independent open coding to the aggregation of code groups.
\item Conducting an evaluation of \diffcoder, yielding valuable insights from user feedback on both the coding and collaboration experiences, which also shed light on the role of LLMs throughout different stages of the CQA process.
\end{enumerate}

\section{Background of Qualitative Analysis}
\label{sec:background_theory}
Qualitative analysis is an important methodology in HCI and social science for interpreting data from interviews, focus groups, observations, and more~\cite{lazar2017research, flick2013sage}. 
The goal of qualitative analysis is to transform unstructured data into detailed insights regarding key aspects of a given situation or phenomenon, addressing researchers' concerns~\cite{lazar2017research}. Commonly employed strategies include Grounded Theory~\cite{flick2013sage} and Thematic Analysis~\cite{maguire2017doing}.

Grounded Theory (GT) is originally formulated by Glaser and Strauss~\cite{glaser2017discovery, flick2013sage}. Its primary objective is to abstract \textit{theoretical conceptions} based on \textit{descriptive} data~\cite{corbin2009basic, bryant2007sage}. A primary approach in GT involves \textit{coding}, specifically assigning \textit{codes} to data segments. These conceptual \textit{codes} act as crucial bridges between descriptive data and theoretical constructs~\cite{bryant2007sage}. In particular, GT coding involves two key phases: initial and focused coding. In initial coding, researchers scrutinize data fragments—words, lines, or incidents—and add codes to them. During focused coding, researchers refine initial codes by testing them against a larger dataset. Throughout, they continuously compare data with both other data and existing codes~\cite{charmaz2014constructing}, in order to build theoretical conceptions or theories.
Similarly, thematic analysis is another method commonly used for analyzing qualitative data, aimed at identifying, analyzing, and elucidating recurring themes within the dataset~\cite{braun2006using, maguire2017doing}.

Several practical frameworks exist for conducting CQA~\cite{hall2005qualitative, bryant2007sage, richards2018practical}. Particularly, Richards et al.~\cite{richards2018practical} have proposed a six-step methodology rooted in GT and thematic analysis. The methodology encompasses the following steps: \numberedphraselowcase{1}{preliminary organization and planning}: An initial team meeting outlines project logistics and sets the overall analysis plan; \numberedphraselowcase{2}{open and axial coding}: Team members use open coding to identify concepts and patterns, followed by axial coding to link these patterns~\cite{corbin2009basic, flick2013sage}; \numberedphraselowcase{3}{development of a preliminary codebook}: One team member reviews the memos and formulates an initial codebook; \numberedphraselowcase{4}{pilot testing the codebook}: After creating the initial codebook, it is tested on new data. Researchers independently code 2-3 transcripts and note codebook issues in their journals; \numberedphraselowcase{5}{final coding process}: The updated codebook is applied to all data, including initially-coded transcripts; and \numberedphraselowcase{6}{review and finalization of the codebook and themes}: After coding all the transcripts, either by consensus or split coding, the team holds a final meeting to finalize the codebook.

Richards et al. also delineate two distinct CQA approaches: \textit{consensus coding} and \textit{split coding}. \textit{Consensus coding} is more rigorous but time-consuming; each coder independently codes the same data and then engages in a team discussion to resolve disagreements and reach consensus. Conversely, \textit{split coding} is quicker but less rigorous, with coders working on separate data sets. This method leans heavily on the clarity established during the preliminary coding phases and pre-defined coding conventions.

Drawing on the six-step CQA framework by Richards et al., \textbf{\diffcoder is designed to streamline crucial stages of the CQA workflow}. It places particular emphasis on the \textbf{consensus coding} approach, ensuring thorough data discussions and complete resolution of disagreements. It also focus on \textbf{inductive qualitative analysis}, wherein both codes and the codebook evolve during the analytical process. This is in contrast to the work by Xiao et al.~\cite{xiao2023supporting}, which prioritizes the use of LLMs to assist deductive coding based on a pre-existing codebook. The overarching aim is to lower the bar for maintaining the rigor and reliability of the inductive CQA process.

\begin{table*}[!htbp]
\caption{Summary of Sources Informing Our Design Goals}
\label{tab:design_goals}
\scalebox{0.85}{\begin{tabular}{|c|c|c|}
\hline
\textbf{Sources} & \textbf{Content} & \textbf{Design Goals (DG)} \\ \hline
\begin{tabular}[c]{@{}l@{}}Step1: Semi-systematic \\ literature review\end{tabular} & \begin{tabular}[c]{@{}c@{}}insights into the key phases of CQA theories, \\ including grounded theory and thematic analysis, \\ detailing inputs, outputs, and practical considerations \\ for each.\end{tabular} & DG1, DG2, DG4, DG5, DG6, DG7 \\ \hline
\begin{tabular}[c]{@{}c@{}}Step2: Examination of \\ prevalent CQA platforms\end{tabular} & \begin{tabular}[c]{@{}c@{}}insights into essential features, pros, and cons \\ of key CQA platforms, such as Atlas.ti Web\footnote{\url{https://atlasti.com/atlas-ti-web}}, \\ MaxQDA Team Cloud\footnote{\url{https://www.maxqda.com/help-mx20/teamwork/can-maxqda-support-teamwork}}, \\ NVivo Collaboration Cloud\footnote{\url{https://lumivero.com/products/collaboration-cloud/}}, \\ and Google Docs. \end{tabular} & DG1, DG2, DG3, DG8 \\ \hline
\begin{tabular}[c]{@{}c@{}}Step3: Preliminary interviews \\ with researchers with \\ qualitative analysis experience \end{tabular} & \begin{tabular}[c]{@{}c@{}}insights into \diffcoder workflow, features, and \\ design scope through expert feedback, concerns, \\ and recommendations.\end{tabular} & DG1, DG4 \\ \hline
\end{tabular}}
\end{table*}

In the following, we delineate the terms and concepts integral to this paper:

\begin{itemize}   
   \item \textbf{Code:} A code is typically a succinct word or phrase created by the researcher to encapsulate and interpret a segment of qualitative data. This facilitates subsequent pattern detection, categorization, and theory building for analytical purposes~\cite{saldana2021coding}.
    \item \textbf{Coding:} Coding serves as a key method for analyzing qualitative data. It involves labeling segments of data with codes that concurrently categorize, encapsulate, and interpret each individual data point~\cite{flick2013sage}.
    \item \textbf{Codebook/Themes/Code groups:} A codebook is a hierarchical collection of code categories or thematic structure, typically featuring first and second-order themes or code groups, definitions, transcript quotations, and criteria for including or excluding quotations~\cite{richards2018practical, saldana2021coding}.
    \item \textbf{Agreement/Consensus:} Agreement or consensus is attained through in-depth discussions among researchers, where divergent viewpoints are scrutinized and potentially reconciled following separate rounds of dialogue~\cite{mcdonald2019reliability}. The degree of agreement among multiple coders serves as an indicator of the analytical rigor of a study~\cite{cornish2013collaborative}.
    \item \textbf{Intercoder Reliability (IRR):} IRR is a numerical metric aimed at quantifying agreement among multiple researchers involved in coding qualitative data~\cite{mcdonald2019reliability, o2020intercoder}.
    \item \textbf{Independence:} Typically, open coding and initial code development are undertaken independently by individual team members to minimize external influence on their initial coding choices~\cite{hall2005qualitative, cornish2013collaborative}.
    \item \textbf{Data units/Unit-of-analysis:} The unit-of-analysis (UoA) specifies the granularity at which text annotations are made, such as at the flexible or sentence level~\cite{rietz2021cody}. 
\end{itemize}
\section{Related Work}

\subsection{Existing Tools for CQA}

Researchers have proposed multiple platforms and approaches to facilitate CQA~\cite{Gilbert2014, drouhard2017aeonium, ganji2018ease}. For instance, Aeonium~\cite{drouhard2017aeonium} assists coders by flagging of uncertain data, highlighting discrepancies, and permitting additional code definitions and coding history checks.  Code Wizard~\cite{ganji2018ease}, an Excel-embedded visualization tool, supports the code merging and analysis process of CQA by allowing coders to aggregate each individual coding table, automatically sort and compare the coded data, calculate IRR, and generate visualization results. Zade et al.~\cite{zade2018conceptualizing} suggest sorting the text according to its ambiguity, allowing coders to concentrate on disagreements and ambiguities to save time and effort. 
The primary goal of these works is to simplify code comparison and identify disagreements and ambiguities~\cite{chen2018using}, thereby enhancing code understanding among coders and streamlining the consensus-building process.

There are also several commercial software (e.g., NVivo\footnote{\url{https://lumivero.com/products/collaboration-cloud/}}, MaxQDA\footnote{\url{https://www.maxqda.com/help-mx20/teamwork/can-maxqda-support-teamwork}}, and Atlas.ti Web\footnote{\url{https://atlasti.com/atlas-ti-web}}) available that support collaborative coding in various ways. For code comparison and discussion, these systems enable users to export and import individually coded documents, facilitating line-by-line, detailed discussions among the coding team for conflict resolution and code consolidation. They also permit coders to add memos for noting concerns and ambiguities to be addressed in discussions. Specifically, Atlas.ti Web (very different from its local version) allows coders to collaborate in real-time within a shared online space for data and code sharing, a feature also present in Google Docs, albeit not tailored for CQA. While it aligns closely with our objective of streamlining CQA workflows by eliminating the need for downloading or uploading documents, they lean towards a "less rigorous" coding method. Given that codes and data are persistently visible to all coders, they do not facilitate "independent" open coding within a coding team. This feature is also available in the latest version of NVivo's collaboration tools\footnote{Accessed on September 13, 2023.}.

Our work on \diffcoder enriches the existing literature by offering a one-stop, end-to-end workflow that seamlessly transitions the output of one stage into the input for the next. It also leverages prior design considerations to aid coders in reaching consensus. Ultimately, our objective with \diffcoder is to streamline the entire CQA process, thereby lowering the barriers to adopting a rigorous approach. This rigor manifests through the inclusion of key CQA stages, the preservation of coder independence, and the fostering of thorough discussions that lead to informed coding decisions based on both agreements and disagreements.

\subsection{AI-assisted (C)QA Systems}

While the utilization of AI to aid in different aspects of the qualitative coding process has garnered increasing interest~\cite{Muller2016, chen2018using, jiang2021supporting, feuston2021putting}, the majority of current research focuses on leveraging AI to aid individual qualitative coding.

Feuston et al.~\cite{feuston2021putting} outlined various ways AI can be beneficial at different stages of QA. For example, AI can provide preliminary insights into large data sets through semantic network analysis before formal inductive coding begins. It can also suggest new codes based on the initial coding work already performed by coders. Particularly relevant to our research with \diffcoder is their emphasis on coding stages: "when, how, and whether" to introduce AI is an important consideration. This aligns with our exploration that AI should, and can, perform different functions and have distinct task allocations at various key stages of CQA.

For system work, Cody~\cite{rietz2021cody} utilizes supervised techniques to enable researchers to define and refine code rules that extend coding to unseen data, while PaTAT~\cite{gebreegziabher2023patat} provides a program synthesizer that learns human coding patterns and serves as a reference for users. Scholastic~\cite{hong2022scholastic} partially shares our goal, specifically aiming to maintain a focused workflow while utilizing codes generated by coders as input for subsequent stages, such as a learning pattern for AI, and as filters to visualize the distribution of emerging knowledge clusters.
On the collaboration side, Gao et al.~\cite{gao2023coaicoder} have identified opportunities to use AI to facilitate CQA efficiency at the early stage of CQA. They contend that utilizing a shared AI model, trained on the coding team's past coding history, can expedite the formation of a shared understanding among coders.

Although these AI-related works utilize traditional AI technologies rather than the latest LLMs, their insights and design considerations have informed the development of our \diffcoder workflow. They prompt us to further consider AI's role throughout the process, as well as the potential advantages and concerns that such assistance might introduce.

\subsection{Using LLMs in Qualitative Analysis}

Recent advancements in LLMs like GPT-3.5 and GPT-4 offer promising text generation, comprehension, and summarization capabilities~\cite{openai2023gpt4}. To enhance coding efficiency, Atlas.ti Web has incorporated OpenAI's GPT models to provide one-click code generation and AI-assisted code suggestions\footnote{\url{https://atlasti.com/atlas-ti-ai-lab-accelerating-innovation-for-data-analysis}}. 
Other software predominantly depend on manual human evaluation or basic AI applications, such as word frequency counting or sentiment analysis.

On the research side, Byun et al.~\cite{byun2023dispensing} posed the question: \textit{"Can a model possess experiences and utilize them to interpret data?"} They examined various prompts to assess theme generation by models such as \textit{text-davinci-003}, a fine-tuned variant, and ChatGPT (referred to as gpt-turbo in their experiment). Their approach involved methods like one-shot prompting and question-only techniques. Their findings suggested that these models are adept at producing human-like themes and posing thought-provoking questions.
Furthermore, they discovered that subtle changes in the prompt — transitioning from \textit{"important"} to \textit{"important HCI-related"} or \textit{"interesting HCI-related"} — yielded more nuanced results.
Additionally, Xiao et al.~\cite{xiao2023supporting} demonstrated the viability of employing GPT-3 in conjunction with an expert-developed codebook for deductive coding. Their findings showcased a notable alignment with expert ratings on certain dimensions. Moreover, the codebook-centered approach surpassed the performance of the example-centered design. They also mentioned that transitioning from a zero-shot to a one-shot scenario profoundly altered the performance metrics of LLMs.

In summary, while research on CQA has explored code comparison and the identification of disagreements in specific phases, as well as the use of AI and LLMs for (semi-)automated qualitative analysis, a comprehensive end-to-end workflow that lowers the barrier for user adherence to the standard CQA workflow remains largely unexplored. Furthermore, the seamless integration of LLMs into this workflow, along with the accompanying benefits and concerns, remains an unexplored avenue. Our proposed workflow, \diffcoder, aims to cover key stages such as independent open coding, iterative discussions, and codebook development. Additionally, we offer insights into the integration of LLMs within this workflow.
\section{Design Goals}
We aim to create a workflow that aligns with standard CQA protocols with a lower adherence barrier, while also integrating LLMs to boost efficiency.

\subsection{Method}
\label{sec:DGs:method}
To achieve our goals, we extracted 8 design goals (DG) for \diffcoder from three primary sources (see Table \ref{tab:design_goals}).

\begin{table*}[!t]
\caption{Participant Demographics in Exploration Interview}
\label{tab:iteration}
\scalebox{0.88}{\begin{tabular}{@{}lllll@{}}
\toprule
No. & Fields of Study & Current Position & QA Software & Years of QA \\ \midrule
P1 & HCI, Ubicomp & Postdoc Researcher & Atlas.ti & 4.5 \\
P2 & HCI, NLP & PhD student & \begin{tabular}[c]{@{}l@{}}Google Sheet/\\ Whiteboard\end{tabular} & 4 \\
P3 & HCI, Health & PhD student & Google Sheet & 4 \\
P4 & HCI, NLP & PhD student & Excel & 1.5 \\
P5 & Software Engineering & PhD student & Google Sheet & 1 \\ \bottomrule
\end{tabular}}
\end{table*}

\paragraph{\added{Step1: Semi-systematic literature review}} We initially reviewed established theories and guidelines on qualitative analysis. \added{Given our precise focus on theories such as Grounded Theory and Thematic Analysis and our emphasis on their particular steps, we used a semi-systematic literature review method \cite{snyder2019literature, mantyla2015rapid}. This method is particularly aimed at identifying key themes relevant to a specific topic while offering an appropriate balance of depth and flexibility.  Our results are incorporated into the background section, as detailed in Section \ref{sec:background_theory}, aiming to establish a robust theoretical foundation for our work. It also assists in delineating the inputs, outputs, and practical considerations for each stage of \diffcoder workflow. This method formulates DG1, DG2, DG4, DG5, DG6, DG7. }

\paragraph{\added{Step2: Examination of prevalent CQA platforms}} The semi-systematic literature review was followed by triangulation with existing qualitative analysis platforms, for which we assessed the current state of design by examining their public documents and official websites (the detailed examination are summarized in Appendix Table \ref{tab:appendix:CQA_software}). \added{This examination enables us to gain insights into the critical features, advantages, and drawbacks of these CQA platforms, such as the dropdown list for the selection of historical codes and the calculation of essential analysis metrics. As a result of this triangulation, we successfully extracted new design goals, DG3 and DG8, and refined the existing DG1 and DG2.}

\paragraph{\added{Step3: Preliminary interviews with researchers with qualitative analysis experience}} \added{Based on the primary understanding of the CQA theories, and the primary version of 8 DGs,} we subsequently developed the primary prototype (see Appendix Figures \ref{fig:appendix:phase1}, \ref{fig:appendix:phase2}, and \ref{fig:appendix:phase3}).
\added{We utilized the initial version of \diffcoder to conduct a pilot interview evaluation with five researchers possessing at least one year of experience in qualitative analysis (refer to Table \ref{tab:iteration}). The aim was to gather expert insights into the workflow, features, and design scope of the theory-driven \diffcoder, thereby refining our design goals and adjusting the prototype's primary features. During the evaluation, the researchers were first introduced to the \diffcoder prototype. Subsequently, they shared their impressions, raised questions, and offered suggestions for enhancements.}
We transcribed their interview audio and did a thematic analysis on the interview transcriptions (see analysis results in Appendix Figure \ref{fig:appendix:expert_interview}) 
and refined two of the design goals (DG1 and DG4) based on their feedback.

\subsection{Results for Design Goals}

\paragraph{\textbf{DG1: Supporting key CQA phases to encourage stricter adherence to standardized CQA processes}}
\label{sec:design_goal_1}
Our primary goal is the creation of a mutually agreed codebook among coders, essentially focusing on the inductive qualitative analysis process. Therefore, from the six-step CQA methodology~\cite{richards2018practical}, we are particularly concerned with "open and axial coding", "iterative discussion", and the "development of a codebook". 

Although complying with CQA steps is critical for deriving robust and trustworthy data interpretations~\cite{richards2018practical}, the existing software workflows and AI integrations are quite demanding. These systems currently do not offer a centralized and focused workflow; there is a noticeable absence of fluidity between stages, where the output of one phase should ideally transition seamlessly into the input of the next. This deficiency complicates the sensemaking process among coders and often discourages them from adhering to the standardized CQA workflow. This sentiment is mirrored by an expert (P1) who remarked, \textit{"In a realistic scenario, how many people do follow this [standard] flow? I don't think most people follow."}

In response, we have tailored a workflow that integrates the key CQA stages we identified. This streamlined process assists the coding team in aligning with the standard coding procedure, ensuring results from one phase transition seamlessly into the next. Our goal is to simplify adherence to the standard workflow by making it more accessible.

\paragraph{\textbf{DG2: Supporting varying levels of coding independence at each CQA stage to ensure a strict workflow.}}
\label{sec:design_goal_2}
In Grounded Theory~\cite{corbin2009basic, richards2018practical}, a primary principle is to enable coders to independently produce codes, cultivate insights from their own viewpoints, and subsequently share these perspectives at later stages. However, we have found that widely-used platforms such as Atlas.ti Web and NVivo, while boasting real-time collaborative coding features, fall short in providing robust support for independent coding. The persistent visibility of all raw data, codes, and quotations to all participants may potentially bias the coding process. Moreover, Gao et al.~\cite{gao2023coaicoder} found that in scenarios prioritizing efficiency, some coders are willing to compromise independence, which could potentially impact coding rigor. 

In response, our workflow designates varying levels of coder independence at different stages: strict separation during the independent open coding phase and mutual code access in the discussion and code grouping phases. We aim to ensure that coders propose codes from their unique perspectives, rather than prematurely integrating others' viewpoints, which could compromise the final coding quality.

\paragraph{\textbf{DG3: Supporting streamlined data management and synchronization within the coding team.}}
\label{sec:design_goal_3}
While Atlas.ti Web has faced criticism for its lack of support for coder independence~\cite{gao2023coaicoder}, as outlined in DG2, it does offer features like \textbf{synchronization} and \textbf{centralized data management}. Through a web-based application, these features allow teams to manage data preprocessing and share projects, ensuring seamless coding synchronization among members. The sole requirement for participation is a web browser and an Atlas.ti Web account.
In contrast, traditional software like MaxQDA and NVivo lack these capabilities. This absence necessitates additional steps, such as locally exporting coding documents post-independent coding and then sharing them with team members\footnote{\url{https://www.maxqda.com/help-mx20/teamwork/can-maxqda-support-teamwork}}. These steps may introduce obstacles to a smooth and focused CQA process. However, as mentioned in DG2, Atlas.ti Web sacrifices coding independence.

In response, we strive to strike a balance between data management convenience and coding independence, facilitating seamless data synchronization and management via a web application while maintaining design features that support independent coding.

\paragraph{\textbf{DG4: Supporting interpretation at the same level among coders for efficient discussion.}}
\label{sec:design_goal_4}
As per Saldana's qualitative coding manual~\cite{saldana2021coding}, coders may use a "splitter" (e.g., line-by-line) or a "lumper" (e.g., paragraph-by-paragraph) approach. This variation can lead coders to work on different levels of granularity, resulting in many extra efforts to align coding units among coders for line-by-line or code-by-code comparison, in order to make them on the same level to determine if they have an agreement or consensus, not to mention the calculation of IRR~\cite{gao2023coaicoder}. Therefore, standardizing and aligning data units for coding among teams is essential to facilitate efficient code comparisons and IRR calculations~\cite{kurasaki2000intercoder, o2020intercoder, ganji2018ease}.
Two prevalent approaches to achieve this are: 1) allowing the initial coder to finish coding before another coder commences work on the same unit~\cite{diaz2023applying, kurasaki2000intercoder, o2020intercoder}, and 2) predefining a fixed text unit for the team, such as sentences, paragraphs, or conceptually significant "chunks"~\cite{o2020intercoder, kurasaki2000intercoder}.

In response, we aim to enhance code comparison efficiency by offering coders predefined coding unit options on \diffcoder, thereby ensuring alignment between their interpretations.
However, it is important to recognize an intrinsic trade-off between \textbf{unit selection flexibility} and \textbf{effort expenditure}. While reduced flexibility can decrease the effort needed to synchronize coders' understanding in discussions, it may also constrain users' freedom in coding.
According to expert feedback, our workflow represents an "ideal" scenario. As one expert (P3) noted, \textit{"I think overall the \diffcoder workflow pretty interesting... However, I think the current workflow is a very perfect scenario. What you haven't considered is that in qualitative coding, there's often a sentence or a section that can be assigned to multiple codes. In your current case, you are assigning an entire section into just one code."}
Additionally, our proposed workflow appears to operate under the assumption that coding is applied to specific, isolated units, failing to account for instances where the same meaning is distributed across different data segments. 
\textit{"Because sometimes [for a code] you need one part of one paragraph, the other part is in another paragraph. right?" }(P1)

\paragraph{\textbf{DG5: Supporting coding assistance with LLMs while preserving user autonomy.}}
\label{sec:design_goal_5}
As Jiang et al.~\cite{jiang2021supporting} suggested, AI should not replace human autonomy, participants in their interview said that \textit{"I don't want AI to pick the good quotes for me..."}. AI should only offer recommendations when requested by the user, after they have manually labeled some codes, and support the identification of overlooked codes based on existing ones. 
To control user autonomy, the commercial software, Atlas.ti Web, has transitioned from auto-highlighting quotations and generating code suggestions via LLMs for all documents with a single click, to now allowing users to request such suggestions on demand \footnote{\url{https://atlasti.com/atlas-ti-ai-lab-accelerating-innovation-for-data-analysis}, accessed on 14th August 2023}. The platform's earlier AI-driven coding, although time-saving, compromised user control in the coding process.

In response, we emphasize user autonomy during the coding process, letting coders first formulate their own codes and turning to LLM assistance only upon request.

\paragraph{\textbf{DG6: Facilitating deeper and higher-quality discussion.}}
\label{sec:design_goal_6}

As highlighted in Section \ref{sec:background_theory}, \diffcoder's primary objective is to foster consensus among coders~\cite{anderson2016all, richards2018practical}. This demands quality discussions rooted in common ground~\cite{patel2012factors, olson2000distance} or shared mental model~\cite{gorman2014team, gorman2010team}.
Common ground pertains to the information that individuals have in common and are aware that others possess~\cite{clark1991grounding, bjorn2014does}. This grounding is achieved when collaborators engage in deep communication~\cite{bjorn2014does}. Similarly, shared mental model is a conception in team coordination theory~\cite{malone1994coordination, entin1999adaptive}. The development of this shared mental model can enable team members to anticipate one another's needs and synchronize team members' efforts, facilitating implicit coordination without the necessity for explicit interaction~\cite{entin1999adaptive}. This becomes particularly valuable in enabling high-quality and efficient coordination, especially when time is limited~\cite{iresearchnet_2016_team_mental_model}. 


In response, we aim to establish common ground or shared mental model among the team to 1) facilitate deeper and higher-quality discussion by surfacing underlying coding disagreements; 2) concentrate coders' efforts on the most critical parts that need the most discussion~\cite{drouhard2017aeonium, zade2018conceptualizing}.

\paragraph{\textbf{DG7: Facilitating cost-effective, fair coding outcomes and engagement via LLMs.}}
\label{sec:design_goal_7}
Once the common ground is established, achieving a coding outcome that is cost-effective, fair, and free from negative effects becomes a challenging yet crucial task~\cite{jameson2003enhancing, emamgholizadeh2022supporting}. To reach a consensus, the team often engages in debates or invests time crafting code expressions that satisfy all coders~\cite{emamgholizadeh2022supporting}, significantly prolonging the discussion. In addition, Jiang et al. ~\cite{jiang2021supporting} reveal that team leaders or senior members may have the final say on the codes, potentially introducing bias.

In response, our objective is to foster deep, efficient, and balanced discussions within the coding team. We ensure that every coder's prior open coding decisions are respected, allowing them to actively participate in both discussions and the final decision-making process, with the support of LLMs.

\paragraph{\textbf{DG8: Enhancing the team's efficiency in code group generation}}
\label{sec:design_goal_8}

Prevalent QA software like Atlas.ti, MaxQDA, and NVivo prominently feature a code manager. This tool lets coders track, modify, and get a holistic view of their current code assignments. It plays a vital role in facilitating discussions, proposing multiple code groups, and aiding code reuse during coding. Meanwhile, Feuston et al.~\cite{feuston2021putting} noted some participants used AI tools to auto-generate final code groups from human-assigned codes. 

In response, we offer the code manager that allows for manual editing and adjustment of code groups. Additionally, we aim to integrate automatic code group generation to streamline the coding process via the assistance of LLMs.

\begin{figure*}[!t]
  \centering
  \includegraphics[width=\textwidth]{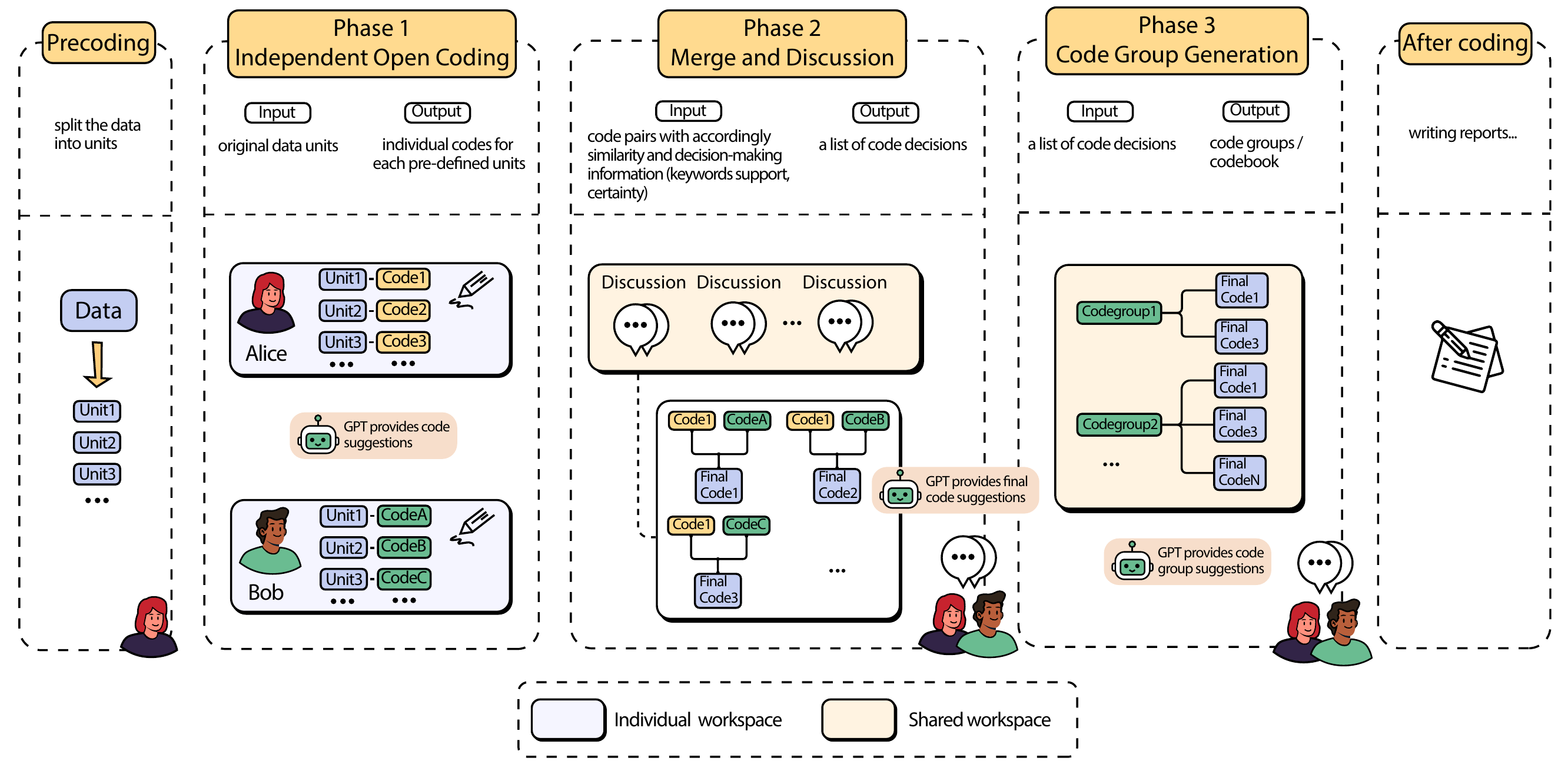}
  \caption{\diffcoder Workflow. The lead coder Alice first splits qualitative data into small units of analysis, e.g., sentence, paragraph, prior to the formal coding. Alice and Bob then: Phase 1: independently perform open coding with GPT assistance; Phase 2: merge, discuss, and make decisions on codes, assisted by GPT; Phase 3: utilize GPT to generate code groups for decided codes and perform editing. They can write reports based on the codebook and the identified themes after the formal coding process.}
  \label{fig:flow}
\end{figure*}

\section{\diffcoder System}
With the aforementioned design goals in mind, we finalized the \diffcoder system and its CQA workflow (refer to Figure \ref{fig:flow}).

\subsection{\diffcoder Workflow \& Usage Scenario}

We introduce an example scenario to demonstrate the usage of \diffcoder (see Figure \ref{fig:interface:editing}). Suppose two coders \textbf{Alice} and \textbf{Bob} are conducting qualitative coding for their data. The lead coder, \textbf{Alice}, first creates a new project on \diffcoder, then imports the data, specifies the level of coding as "paragraph", and invites \textbf{Bob} to join the project. After clicking on {\large \MakeUppercase{c}\textsc{reate project}}, \diffcoder's parser will split the imported raw data into units (paragraph in this case). The project can then be shown on both coders' interfaces.

\subsubsection{Phase 1: Independent Open Coding} In Phase 1, \textbf{Alice} and \textbf{Bob} individually formulate codes for each unit in their separate workspaces via the same interface. Their work is done independently, with no visibility into each other's codes.
If \textbf{Alice} wants to propose a code for a sentence describing a business book for students, she can either \textit{craft} her own code, choose from code recommendations generated by the GPT model (e.g., \texttt{"Excellent guide for new college students"}, \texttt{"Insightful read on business fundamentals"}, \texttt{"How A Business Works": semester's gem}), or \textit{pick} one of the top three most relevant codes discovered by GPT in her coding history, and making modifications as needed. 
She can then \textit{select} relevant keywords/phrases (e.g., \texttt{"excellent book"}, \texttt{"college student"}) from the 
{\large \MakeUppercase{R}\textsc{aw data}}
cell that supports her proposed code, which will be added to the 
{\large \MakeUppercase{K}\textsc{eywords support}} beside her proposed code. She can also \textit{assign} a {\large\MakeUppercase{C}\textsc{ertainty}}, ranging from 1 to 5, to the code. This newly generated code will be included in \textbf{Alice}'s personal {\large \MakeUppercase{C}\textsc{odebook}} and can be \textit{viewed} by her at any time. 
They can \textit{check} the progress of each other in the {\large \MakeUppercase{P}\textsc{rogress}} at any time (see Figure~\ref{fig:interface:compare}).

\begin{figure*}[!t]
    \centering
    \includegraphics[width=\linewidth]{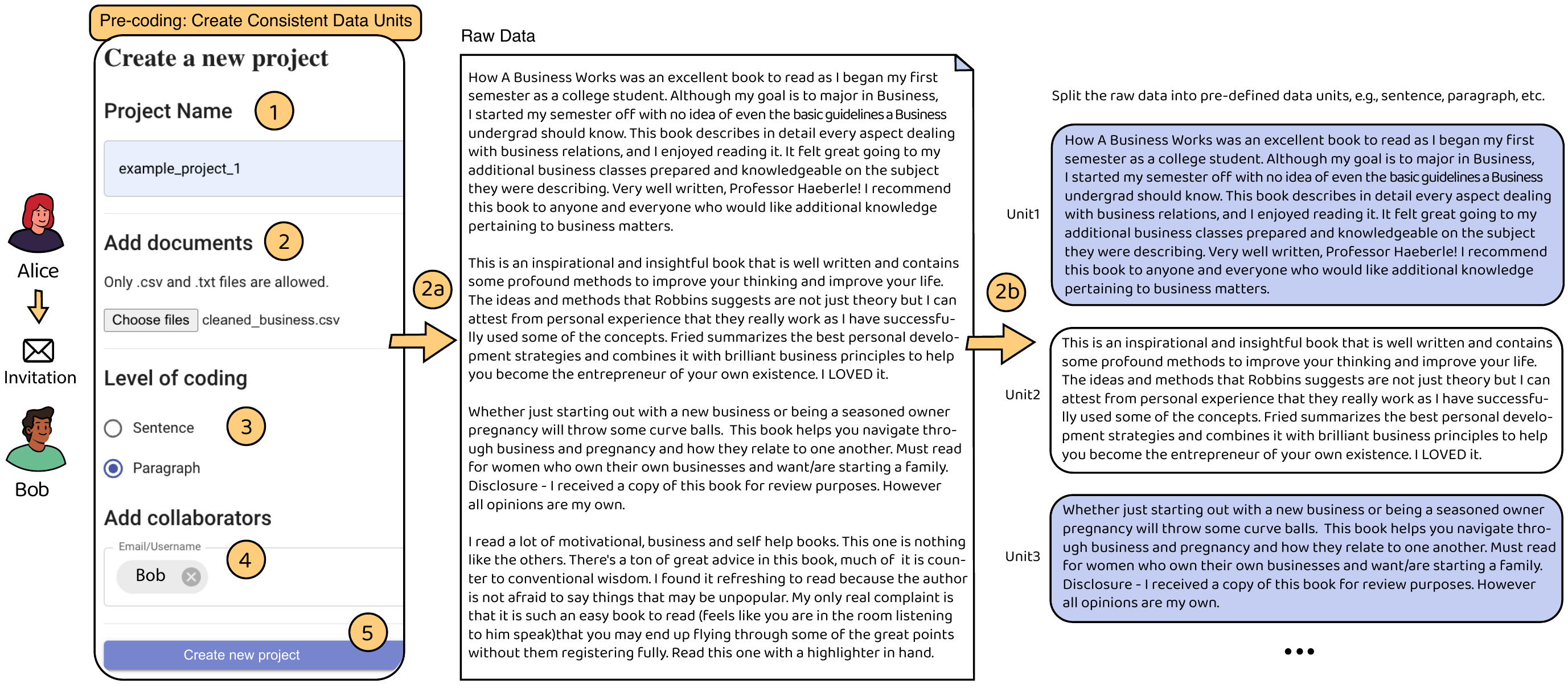}
    \caption{Precoding: establish consistent data units and enlist coding team during project creation. The primary coder, Alice, can: 1) name the project, 2) incorporate data, ensuring it aligns with mutually agreed data units, 2a) illustrate how \diffcoder manages the imported data units, 3) define the coding granularity (e.g., sentence or paragraph), 4) invite a secondary coder, Bob, to the project, and 5) initiate the project.}
    \label{fig:interface:newproject}
\end{figure*}

\subsubsection{Phase 2: Code Merging and Discussion} 
\begin{figure*}[!t]
   \centering
  \includegraphics[width=\textwidth]{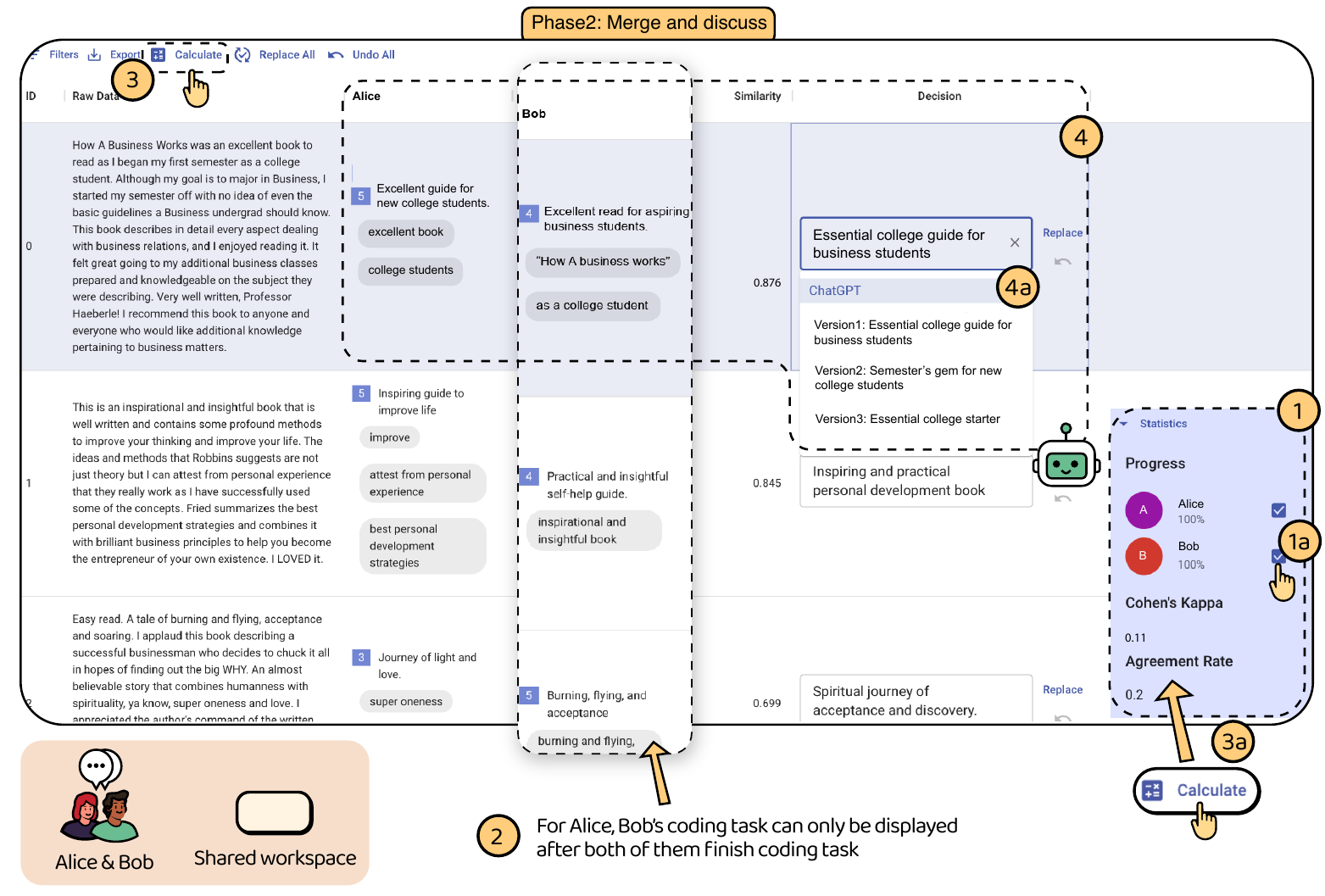}
  \caption{Comparison Interface for Phase 2. Users can discuss and reach a consensus by following these steps: 1) reviewing another coder's progress and 1a) clicking on the checkbox \textbf{only if} both individuals complete their coding, 2) two coders' codes are listed in the same interface, 3) calculating the similarity between code pairs and 3a) IRR between coders, 4) sorting the similarity scores from highest to lowest and identifying (dis)agreements, and 4a) making a decision through discussion based on the initial codes, raw data, and code supports or utilizing the GPT's three potential code decision suggestions. Additionally, users have the option to "Replace" the original codes proposed by two coders and revert back to the original codes if required. They can also replace or revert all code decisions with a single click on the top bar. }
  \label{fig:interface:compare}
\end{figure*}

Figure \ref{fig:interface:compare} depicts the shared workspace where coding teams collaborate, discussing their code choices and making final decisions regarding the codes identified in Phase 1.
After completing coding, \textbf{Alice} can \textit{check} the {\large 
 \MakeUppercase{C}\textsc{heckbox}} next to \textbf{Bob}'s name once she sees that his progress is at 100\%. 
Subsequently, she can \textit{click} the {\large 
 \MakeUppercase{C}\textsc{alculate}} button to generate quantitative metrics such as similarity scores and IRR (Cohen's Kappa and Agreement Rate\footnote{The calculation methods differ between these two metrics. Cohen's kappa is a more intricate method for measuring agreement, as detailed in ~\cite{mchugh2012interrater}. On the other hand, the Agreement Rate represents the percentage of data on which coders concur.}) within the team. The rows are then sorted by similarity scores in descending order.
 
\textbf{Alice} can then share her screen via a Zoom meeting with \textbf{Bob} to {\large 
 \MakeUppercase{C}\textsc{ompare and discuss}} their codes, starting from code pairs with high similarity scores. 
For instance, \textbf{Alice}'s code \texttt{"Excellent guide for new college students"} with a certainty of 5 includes \texttt{"excellent book"} and \texttt{"college student"} supports, while \textbf{Bob}'s code \texttt{"Excellent read for aspiring business students"} with a certainty of 4 includes \texttt{“How A business works”} and \texttt{"as a college student"} as {\large 
 \MakeUppercase{K}\textsc{eywords support}}. 
The similarity score between their codes could be 0.876 (close to 1), showing a high agreement. During the discussion, they both agreed that the final code should contain the word "student" due to their similar {\large 
 \MakeUppercase{K}\textsc{eywords support}}, but they cannot reach a consensus about the final expression of the code, they then seek GPT suggestions (e.g., \texttt{"Essential college guide for business students"}, \texttt{"Semester's gem for new college students"}, \texttt{Essential college starter}), and decide the final code decision for this unit is \texttt{"Essential college guide for business students"}. 
However, if the code pair presents a low similarity score, they must allocate additional time to scrutinize the code decision information and identify the keywords that led to different interpretations.

Once all code decisions have been made, \textbf{Alice} can then \textit{click} on {\large 
 \MakeUppercase{R}\textsc{eplace}} to replace the original codes, resulting in an update of Cohen's Kappa and Agreement Rate. This action can be undone by clicking on {\large 
 \MakeUppercase{U}\textsc{ndo}}.

\begin{figure*}[!t]
    \centering
    \includegraphics[width=\linewidth]{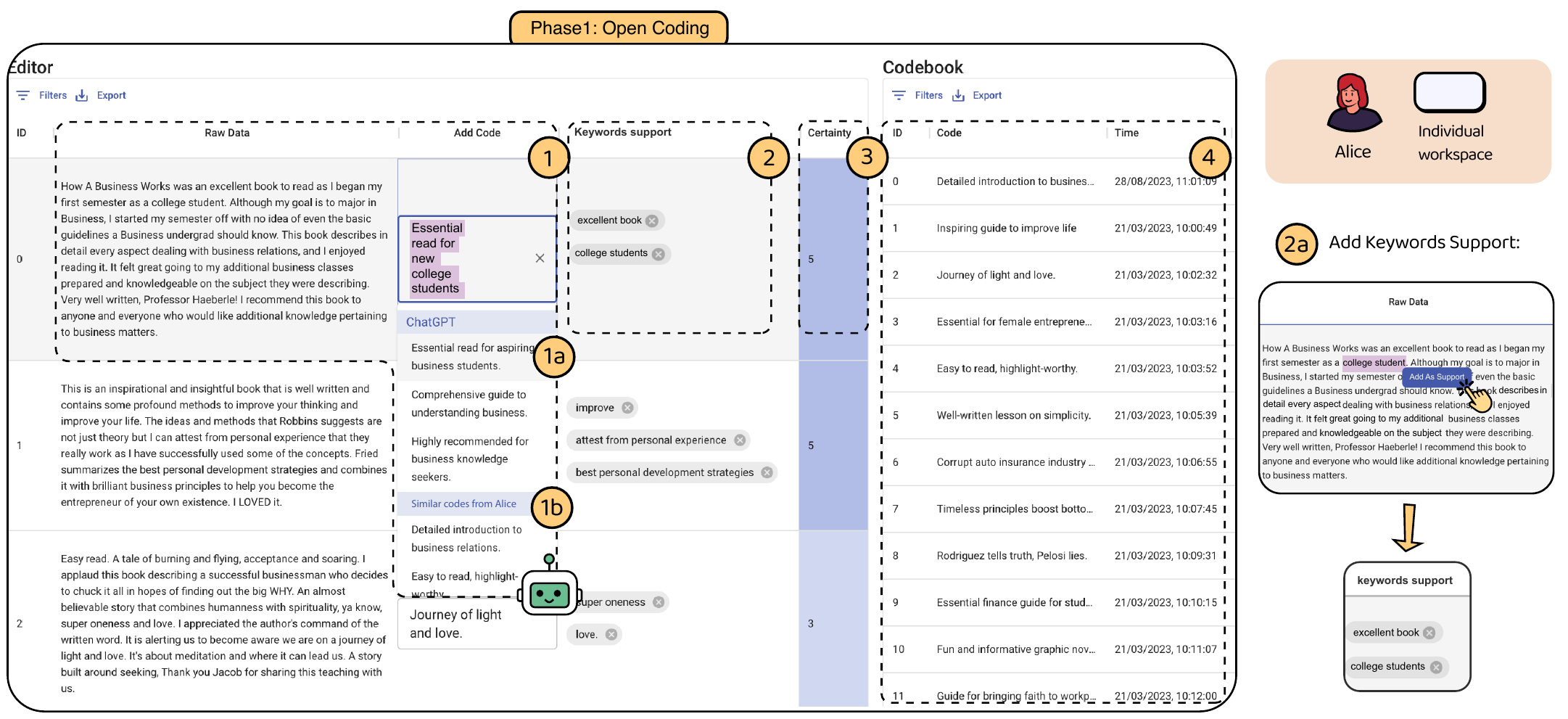}
    \caption{Editing Interface for Phase 1: 1) inputting customized code for the text in "Raw Data", either 1a) choosing from the GPT's recommendations, 1b) choosing from the top three relevant codes; 2) adding keywords support by 2a) selecting from raw data and "Add As Support"; 3) assigning a certainty level ranging from 1 to 5, where 1="very uncertain" and 5="very certain"; and 4) reviewing and modifying the individual codebook.}
    \label{fig:interface:editing}
\end{figure*}

\subsubsection{Phase 3: Code Group Generation} 
\begin{figure*}[!htbp]
    \centering
    \includegraphics[width=\textwidth]{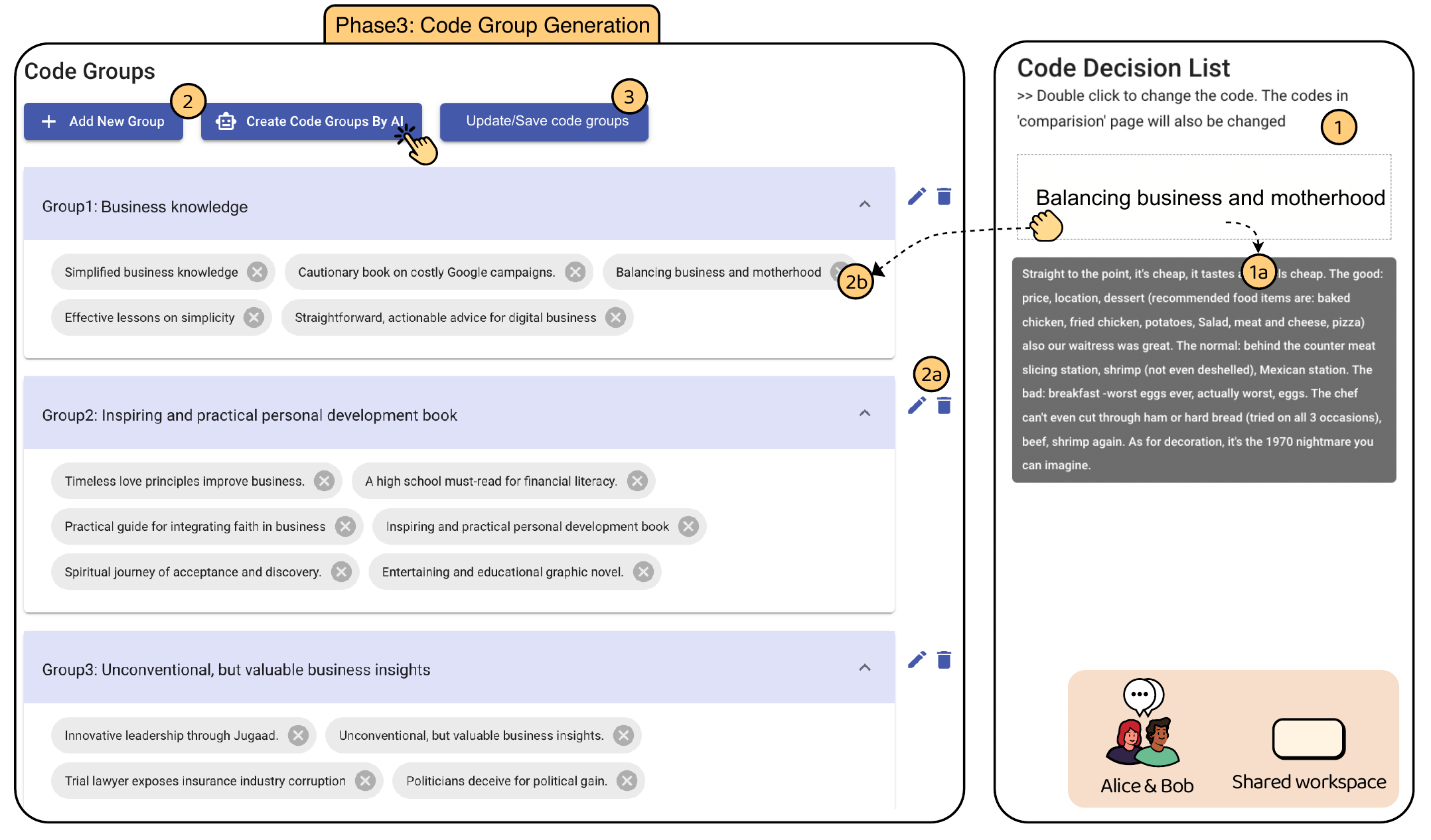}
    \caption{Code Group Interface. It enables users to manage their code decisions in a few steps: 1) the code decisions are automatically compiled into a list of unique codes that users can edit by double-clicking and accessing the original data by hovering over the code. 2) users can group their code decisions by using either "Add New Group" or "Create Code Groups By AI" options. They can then 2a) name or delete a code group or use AI-generated themes, and 2b) drag the code decisions into code groups. 3) Finally, users can save and update the code groups.}
    \label{fig:interface:codegroup}
\end{figure*}

Once \textbf{Alice} and \textbf{Bob} have agreed on the final code decisions for all the units, the code decision list will be displayed on the code grouping interface, as shown in Figure~\ref{fig:interface:codegroup}. This interface is shared uniformly among the coding team. 
For further discussion, \textbf{Alice} can continue to share her screen with \textbf{Bob} on Zoom. She can \textit{hover over} each {\large\MakeUppercase{C}\textsc{ode Decision}} to refer to the corresponding raw data or \textit{double-click} to edit.
\textbf{Alice} and \textbf{Bob} can collaborate to propose the final code groups by \textit{clicking} on {\large 
 \MakeUppercase{A}\textsc{dd new group}} and \textit{drag} the code decisions into the new code group. For instance, a group \texttt{"Business knowledge"} can include \texttt{"Simplified business knowledge"}, \texttt{"Cautionary book on costly Google campaigns"} and others.
Alternatively, they can request GPT assistance by \textit{clicking} on the {\large 
 \MakeUppercase{C}\textsc{reate code groups by AI}} button to automatically generate several code groups and place the individual code decisions into them. These groups can still be manually adjusted by coders. 
Once they finish grouping, they can proceed to report their findings as necessary.

\subsection{Key Features}

\subsubsection{Three-phase Interfaces}
In alignment with \hyperref[sec:design_goal_1]{DG1}, our objective was to incorporate a workflow that supports the three key phases of the CQA process, as derived from established theories. Accordingly, our system is segmented into three distinct interfaces:
\begin{enumerate}
\item Editing Interface for Phase 1: Independent Open Coding (Figure \ref{fig:interface:editing}).
\item Comparison Interface for Phase 2: Merge and Discuss (Figure \ref{fig:interface:compare}).
\item Code Group Interface for Phase 3: Code Groups Generation (Figure \ref{fig:interface:codegroup}).
\end{enumerate}

\subsubsection{Individual Workspace vs. Shared Workspace}
\label{sec:workspace}

Aligned with \hyperref[sec:design_goal_2]{DG2}, we aim to mirror the distinct levels of independence intrinsic to the CQA process, reflecting the principles of qualitative analysis theories. \diffcoder introduces an \textit{"individual workspace"} — the Editing Interface — allowing users to code individually during the initial phase without visibility of others' coding.
Additionally, for facilitating Phase 2 discussions, \diffcoder unveils a "shared workspace." Here, the checkbox next to each coder's name activates only after both participants complete their individual coding, represented as percentages (0-100\%). This shared interface enables the team to collectively review and discuss coding data within an integrated environment.

\subsubsection{Web-based Platform}

In alignment with \hyperref[sec:design_goal_3]{DG3}, our goal is to harness the synchronization benefits of Atlas.ti Web while preserving the essential independence required for the CQA process. \diffcoder addresses this by using a web-based platform. Here, the lead coder creates a project and invites collaborators to engage with the same project. As outlined in section \ref{sec:workspace}, upon the completion of individual coding, participants can effortlessly view the results of others, eliminating the need for downloads, imports, or further steps.

\subsubsection{Consistent Data Units for All Users}

Aligned with \hyperref[sec:design_goal_4]{DG4}, our objective is to synchronize coders' interpretation levels to boost discussion efficiency. \diffcoder facilitates this by segmenting data into uniform units (e.g., sentences or paragraphs) that are collaboratively determined by all coders prior to data importation or the onset of coding task.

\subsubsection{LLMs-generated Coding Suggestions Once the User Requests}

Aligned with \hyperref[sec:design_goal_5]{DG5}, we aim to empower coders to initially develop their own codes and then seek LLMs' assistance when necessary, striking a balance between user autonomy and the advantages of LLMs' support. Apart from proposing their own codes, \diffcoder offers LLMs-generated code suggestions when a user interacts with the input cell. These suggestions appear in a dropdown list for the chosen data unit after a brief delay (\added{$\approx$5 seconds}\footnote{\added{We established a default 2-second delay alongside GPT API's approximate 3-second delay. Investigating the optimal delay is beyond our current research scope, we acknowledge this as a limitation and plan to address it in future research.}}), allowing users time to think about their own codes first. 
At the same time, \diffcoder identifies and provides the three most relevant codes from the current individual codebook for the given text unit, ensuring coding consistency when reusing established codes.

\subsubsection{A Shared Workspace for Deeper Discussion}

In alignment with \hyperref[sec:design_goal_6]{DG6}, our goal is to establish a shared understanding and foster richer, more substantive discussions. \diffcoder supports this goal through three key features.

\begin{enumerate}
    
    \item \textit{Documenting Decision-making Rationale.} In Phase 1, \diffcoder allows users to select keywords, phrases, and their coding certainty as supporting evidence. These highlighted elements can represent pivotal factors influencing the user's coding decision. \diffcoder further facilitates users in rating their certainty for each code on a scale from 1 (least certain) to 5 (most certain) to mark the ambiguity.

    \item \textit{Side-by-Side Comparison in A Shared Workspace}. Building on \hyperref[sec:design_goal_6]{DG6}'s emphasis on establishing common ground, \diffcoder presents all users' coding information for the relevant data units side-by-side. This display includes the original data units, supporting keywords, and indicators of labeled certainty scores. This layout facilitates direct comparison and nuanced discussions.

    \item \textit{Identifying (Dis)agreements.} \diffcoder simplifies the process of spotting (dis)agreements by calculating the Similarity of the code pair of each unit. This analysis can be executed in 3-10 seconds for all data units. Similarity scores for code pairs range from 0 (low similarity) to 1 (high similarity). For ease of interpretation, these scores can be sorted in descending order, with higher scores indicating stronger agreements.
\end{enumerate}

\subsubsection{LLMs as a Group Recommender System}
In alignment with \hyperref[sec:design_goal_7]{DG7}, our aim is to foster cost-effective and equitable coding outcomes utilizing LLMs. \diffcoder achieves this by serving as an LLM-based group recommender system~\cite{jameson2003enhancing}: when users struggle to finalize a code, \diffcoder proposes three code decision suggestions specific to the code pair, taking into account the raw data, codes from each user, keywords support, and certainty scores. Users can then select and customize these suggestions to reach a conclusive coding decision.

\subsubsection{Formation of LLMs-based Code Groups}
Consistent with \hyperref[sec:design_goal_8]{DG8}, our objective is to optimize the process of code group creation to enhance efficiency. To this end, \diffcoder introduces the Code Group interface to provide two key functions:

\begin{enumerate}
    \item \textit{Accessing Original Data via the Final Code Decision List.} \diffcoder streamlines final code decisions, presenting them on the right-hand side of the interface. Hovering over a code reveals its originating raw data. Additionally, by double-clicking on an item within the code decision list, users can amend it, and the corresponding codes are updated accordingly.
    
    \item \textit{Managing Code Groups.} With \diffcoder, users can effortlessly craft, rename, or delete code groups. They can drag codes from the decision list to a designated code group or remove them. To save users the effort of building groups from scratch, \diffcoder provides an option to enlist GPT's help in organizing code decisions into preliminary groupings. This offers a foundation that users can then adjust, rename, or modify.
\end{enumerate}

\subsection{Prompts Design}
\diffcoder leverages OpenAI's ChatGPT model (gpt-3.5-turbo)\footnote{\url{https://platform.openai.com/docs/models/gpt-3-5}} to provide code and code group suggestions. \added{Throughout the three phases, GPT is tasked with the role of "a helpful qualitative analysis assistant", aiding researchers in the development of codes, code decisions, and primary code groups that are crucial for subsequent stages. Additionally, we have tailored different prompts for distinct types of codes. For instance, we use "descriptive codes for raw data" and "relevant codes derived from coding history" (in Phase 1), ensuring a tailored approach for each coding requirement.
The prompts, along with the users' text, are simultaneously sent to GPT for processing.}
All prompts used are listed in Appendix Table \ref{tab:system:prompt1}, \ref{tab:system:prompt2} and \ref{tab:system:prompt3}. To ensure code suggestions have diversity without being overly random, the temperature parameter is set at 0.7.

\subsection{System Implementation}
\subsubsection{Web Application}
The front-end implementation makes use of the react-mui library\footnote{\url{https://mui.com/}}. Specifically, we employed the DataGrid component\footnote{\url{https://mui.com/x/react-data-grid/}} to construct tables in both the "Edit" and "Compare" interfaces, allowing users to input and compare codes. These tables auto-save user changes through HTTP requests to the backend, storing data in the database to synchronize progress among collaborators. For each data unit, users have their own code, keyword supports, certainty levels, and codebook in the Edit interface, while sharing decisions in the "Compare" interface and code groups in the "Codebook" interface. To prevent users from viewing collaborators' codes before editing is complete, we restrict access to other coders' codes and only show everyone's progress in the "Compare" interface.  
We also utilized the foldable Accordion component\footnote{\url{https://mui.com/material-ui/react-accordion/}} to efficiently display code group lists in the "Codebook" interface, where users can edit, drag and drop decision objects to modify their code groups.
The backend leverages the Express framework, facilitating communication between the frontend and MongoDB. It also manages API calls to the GPT-3.5 model and uses Python to calculate statistics such as similarities.

\subsubsection{Data Pre-processing} We partitioned raw data from CSV and txt files into data units during the pre-processing phase. At the sentence level, we segmented the text using common sentence delimiters such as ".", "...", "!", and "?". At the paragraph level, we split the text using \verb|\n\n|.

\subsubsection{Semantic Similarity and IRR}

In \diffcoder, the IRR is measured using Cohen's Kappa\footnote{Cohen's Kappa is a statistical measure used to evaluate the IRR between two or more raters, which takes into account the possibility of agreement occurring by chance, thus providing a more accurate representation of agreement than simply calculating the percentage of agreement between the raters.} and Agreement Rate. To calculate Cohen's Kappa, we used the "cohen\_kappa\_score" method from scikit-learn package backend\footnote{\url{https://scikit-learn.org/stable/modules/generated/sklearn.metrics.cohen_kappa_score.html}}. Cohen's Kappa is a score between -1 (total disagreement) and +1 (total agreement).
Subsequently, we calculate the Agreement Rate as a score between 0 and 1, by determining the percentage of code pairs whose similarity score exceeds 0.8, indicating that the two coders agree on the code segment. \added{We utilize the \textit{semantic textual similarity} function\footnote{\url{https://www.sbert.net/docs/usage/semantic_textual_similarity.html}} in the \textit{sentence-transformers} package\footnote{\url{https://www.sbert.net/}} to assess agreements and disagreements in coding. This function calculates the semantic similarity between each code pair from two coders (e.g., Alice: \texttt{Excellent guide for new college students} vs. Bob: \texttt{Excellent read for aspiring business students}) for each data unit. A high similarity score (close to 1) indicates agreement between coders, while a low score (close to 0) suggests disagreement.}

\section{User Evaluation}
\added{To evaluate \diffcoder and answer our research questions}, we conducted a within-subject user study involving 16 (8 pairs) participants who used two platforms: \diffcoder and Atlas.ti Web, for qualitative coding on two sets of qualitative data. 



\subsection{Participants and Ethics}
We invited 16 participants with varying qualitative analysis experiences via public channels and university email lists. We involve both experts and non-experts as lowering the bar is particularly important for newcomers or early researchers who might confront significant challenges in adhering to such rigorous workflow \cite{richards2018practical, cornish2013collaborative}.
Among them, 2/16 participants identified as experts, 3/16 considered themselves intermediate, 4/16 as beginners, and 7/16 had no qualitative analysis experience (see details in Appendix Table \ref{tab:appendix:participants}). Participants were randomly matched, leading to the formation of 8 pairs (see Table \ref{tab:result}).
Each participant received compensation of approximately \$22.3 USD for their participation, based on the total duration. The study protocol and the financial compensation based on the duration at the hourly rate was approved by our local IRB.

\subsection{Datasets}
We established two criteria to select the datasets used for the coding task: 1) the datasets should not require domain-specific knowledge for coding, and 2) coders should be able to derive a theme tree and provide insights iteratively. Accordingly, two datasets containing book reviews on "Business" and "History" topics from the \texttt{Books\_v1\_00} category of \texttt{amazon\_us\_reviews} dataset\footnote{\url{https://huggingface.co/datasets/amazon\_us\_reviews/viewer/Books\_v1\_00/train}} were selected. For each of them, we filtered 15 reviews to include only those with a character count between 400 and 700 and removed odd symbols such as \textbackslash{} and <br />. The workload was determined through pilot tests with some participants.

\subsection{Conditions}
\begin{itemize}
    \item \textbf{Atlas.ti Web}: a powerful platform for qualitative analysis that enables users to invite other coders to collaborate by adding, editing, and deleting codes. It also allows for merging codes and generating code groups manually. 
    \item \textbf{\diffcoder}: the formal version of our full-featured platform.
\end{itemize}
The presentation order of both platforms and materials was counter-balanced across participants using a Latin-square design~\cite{lazar2017research}.

\subsection{Procedure}
Each study was conducted virtually via Zoom and lasted around 2 to 3 hours. It consisted of a pre-study questionnaire, training for novice participants, two qualitative coding sessions with different conditional systems, a post-study questionnaire, and a semi-structured interview.

\subsubsection{Introduction to the Task} After obtaining consent, we introduced the task to the pairs of participants, which involved analyzing reviews and coding them to obtain meaningful insights. We introduced research questions they should take into account when coding, such as recurring themes or topics, common positive and negative comments or opinions. We provided guidelines to ensure that the coding was consistent across all participants. Participants could use codes up to 10 words long, \added{add similar codes in one cell per data unit, and include both descriptive and in-vivo codes.}

\subsubsection{Specific Process} Following the introduction, we provided a video tutorial on how to use the platform for qualitative coding. Participants first did independent coding, and then discussed the codes they had found and made final decisions for each unit, ultimately forming thematic groups. \added{We advise coders to first gain a thorough understanding of the text, then seek suggestions from GPT, engage in comprehensive discussions, and finally present code groups that effectively capture the valuable insights they have acquired.}
To ensure they understood the study purpose better, participants were shown sample code groups as a reference for the type of insights they should aim to obtain from their coding. After completing the coding for all sessions, participants were asked to complete a survey, which included a 5-level Likert Scale to rate the effectiveness of the two platforms, and self-reported feelings towards them.

\subsubsection{Data Recording} During the process, we asked participants to share their screens and obtained their consent to record the meeting video for the entire experiment. Once the coding sessions were completed, participants were invited to participate in a post-study semi-structured interview.

\subsubsection{\added{Data analysis}}
We analyzed interview transcripts and observation notes (see Table \ref{tab:observation_notes_1} and \ref{tab:observation_notes_2}) using thematic analysis as described in Braun and Clarke’s methodology~\cite{braun2006using}. After familiarizing ourselves with data and generating initial codes, we grouped the transcripts into common themes based on the content. Next, we discussed, interpreted, and resolved discrepancies or conflicts during the grouping process. Finally, we reviewed the transcripts to extract specific quotes relevant to each theme. We summarized the following key findings.
\section{Results}

\subsection{\added{RQ1: Can CollabCoder support qualitative coders to conduct CQA effectively?}}
\label{sec:results:rq1}

\subsubsection{Key Findings (KF) on features \added{that support CQA}}

\paragraph{\textbf{KF1: \textit{\diffcoder} workflow simplifies the learning curve for CQA and ensures coding independence in the initial stages.}}
\label{sec:results:KF1}

Overall, users found \diffcoder to be better as it supports side-by-side comparison of data, which makes the coding and discussion process \textbf{easier to understand (P2), more straightforward (P7), and beginner-friendly (P4)} than Atlas.ti Web, and P4 noted that \diffcoder had a \textbf{lower learning curve}.

Moreover, \diffcoder workflow preserves coding independence. Experienced users (P11 and P14), familiar with qualitative analysis, find \textit{\diffcoder}'s independent coding feature to be particularly beneficial: \textit{"So you don't see what the other person is coding until like both of you are done. So it doesn't like to affect your own individual coding...[For Atlas.ti Web] the fact like you can see both persons’ codes and I think I'm able to edit the other person's codes as well, which I think might not be very a good practice."}
Similarly, P14 indicated: \textit{"I think \diffcoder is better if you aim for independent coding."}

\paragraph{\textbf{KF2: Individual workspace with GPT assistance is valued for reducing cognitive burden \added{in Phase 1.}}}
\label{sec:results:KF2}

\diffcoder makes it easier for beginner users to propose and edit codes compared to Atlas.ti Web. 7/16 participants appreciated that GPT's additional assistance (P7, P15), which gave them reference (P1) and decreased thinking (P9). Such feelings are predominantly reported by individuals who are either beginners or lack prior experience in qualitative analysis.
As P13 said, \textit{"I think the \diffcoder one is definitely more intuitive in a sense, because it provides some suggestion, you might not use it, but at least some basic suggestions, whereas the Atlas.ti one, you have to take from scratch and it takes more mental load."} (P13).

Some of these beginners also showed displeasure towards GPT, largely stemming from its content summarization level, which users cannot regulate. 
P1 (beginner) found that \textbf{in certain instances, \diffcoder generated highly detailed summaries} which might not be well-suited to their requirements, leading them to prefer crafting their own summaries: \textit{"One is that its summary will be very detailed, and in this case, I might not use its result, but I would try to summarize [the summary] myself."} 
This caused them to question AI's precision and appropriateness for high-level analysis, especially in the context of oral interviews or focus groups.

In addition, when adding codes, our participants indicated that they preferred \textbf{reading the raw data first before looking at the suggestions}, as they believed that reading the suggestions first could influence their thinking process (P1, P3, P4, P14) and introduce bias into their coding: \textit{"So I read the text at first. it makes more sense, because like, if you were to solely base your coding on [the AI agent], sometimes its suggestions and my interpretation are different. So it might be a bit off, whereas if you were to read the text, you get the full idea as to what the review is actually talking about. The suggestion functions as a confirmation of my understanding."} (P4)

\paragraph{\textbf{KF3: Pre-defined data units, documented decision-making mechanisms, and progress bar features collectively enhance mutual understanding \added{in Phase 1 and Phase 2}.}}
\label{sec:results:KF3}

Regarding collaboration, users found that having a pre-defined unit of analysis enabled them to more easily understand the context: \textit{"I am able to see your quotations. Basically what they coded is just the entire unit. But you see if they were to code the reviews based on sentences, I wouldn't actually do the hard work based on which sentence he highlighted. But for \diffcoder, I am able to see at a glance, the exact quotations that they did. So it gives me a better sense of how their codes came about."} (P3) 
Moreover, users emphasized the importance of not only having the quotation but also keeping its context using pre-defined data units, as they often preferred to refer back to the original text. This is because understanding the context is crucial for accurate data interpretation and discussion: \textit{"I guess, it is because like we're used to reading a full text and we know like the context rather than if we were to read like short extracts from the text. the context is not fully there from just one or two line [quotations]."} (P9)

Users also appreciated \diffcoder's keywords-support function, as it \textbf{aided them in capturing finer details (P9)} and facilitated a deeper understanding of the codes added: \textit{"It presents a clearer view about that paragraph. And then it helps us to get a better idea of what the actual correct code should be. But since the other one [Atlas.ti Web] is [...] a little bit more like superficial, because it's based solely on two descriptive words."} (P14)

The progress bar feature in \diffcoder was seen as helpful when collaborating with others. It allowed them to \textbf{manage their time better and track the progress of each coder.} \textit{"I actually like the progress bar because like that I know where my collaborators are."} (P8) Additionally, it acted as a \textbf{tracker to notify the user if they missed out on a part}, which can help to avoid errors and improve the quality of coding. \textit{"So if say, for example, I missed out one of the codes then or say his percentage is at 95\% or something like that, then we will know that we missed out some parts"} (P3)

All the above features collectively improve the mutual understanding between coders, which can decrease the effort devoted to revisiting the original data and recalling their decision-making processes, and deepen discussions in a limited time.

\paragraph{\textbf{KF4: The shared workspace with metrics allows coders to understand disagreements and initiate discussions better \added{in Phase 2}.}}
\label{sec:results:KF4}

In terms of statistics during the collaboration, the similarity calculation and ranking features enable users to \textbf{quickly identify (dis)agreements (P2, P3, P7, P10, P14) to ensure they focus more (P4).} As P14 said, \textit{"I think it's definitely a good thing [to calculate similarity]. From there, I think we can decide whether it's really a disagreement on whether it's actually two different information captured in the two different codes."}
Moreover, the identification of disagreements is reported to \textbf{pave the way for discussion (P1, P8)}: \textit{"So I think in that sense, it just opens up the door for the discussion compared to Atlas.ti...[and]better in idea generation stands and opening up the door for discussion."} (P8) In contrast, Atlas.ti necessitated more discussion initiation on the part of users. 

Nevertheless, ranking similarity using \diffcoder might have a negative effect, as it may make coders focus more on improving their agreements instead of providing a more comprehensive data interpretation: \textit{"I think pros and cons. because you will feel like there's a need to get high similarity on every code, but it might just be different codes. So there might be a misinterpretation."} (P7)

The participants had mixed opinions regarding the usefulness of IRR in the coding process. P9 found Cohen's kappa useful for their report as they do not need to calculate manually: \textit{"I think it's good to have Cohen’s Kappa, because we don't have to manually calculate it, and it is very important for our report. "} However, P6 did not consider the statistics to be crucial in their personal research as they usually do coding for interview transcripts. \textit{"Honestly, it doesn't really matter to me because in my own personal research, we don't really calculate. Even if we have disagreements, we just solve it out. So I can't comment on whether the statistics are relevant, right from my own personal experience."} (P6)

\paragraph{\textbf{KF5: \added{The GPT-generated primary code groups in Phase 3 enable coders to have a reference instead of starting from scratch, thereby reducing cognitive burden. \deleted{KF5: \diffcoder{} facilitates a top-down approach to generate code groups, which decreases the cognitive burden.}}}} 
\label{sec:results:KF5}

Participants expressed a preference for the automatic grouping function of \diffcoder, as it was \textbf{more efficient (P1, P2, P8, P14) and less labor-intensive (P3)}, compared to the more manual approach in Atlas.ti Web. 
In particular, P14 characterized the primary distinction between the two platforms as Atlas.ti Web adopts a "bottom-up approach" while \textbf{\diffcoder employs a "top-down approach".} \added{In this context, the "top-down approach" refers to the development of "overall categories/code groups" derived from the coding decisions made in Phase 2, facilitated by GPT. This approach allows users to modify and refine elements within an established primary structure or framework, thereby eliminating the need to start from scratch. Conversely, the "bottom-up approach" means generating code groups from an existing list, through a process of reviewing, merging, and grouping codes with similar meanings.}
This difference impacts the mental effort required to create categories and organize codes. \textit{"I think it's different also because  Atlas.ti is more like a bottom-top approach. So we need to see through the primary codes to create the larger categories which might be a bit more tedious, because usually, they are the primary codes. So it's very hard to see an overview of everything at once. So it takes a lot of mental effort, but for \diffcoder, it is like a top-down approach. So they [AI] create the overall categories. And then from there, you can edit and then like shift things around which helps a lot. So I also prefer \diffcoder."} (P14) P1 also highlighted that this is particularly helpful when dealing with large amounts of codes, as manually grouping them one-by-one becomes nearly unfeasible.

\begin{figure*}[!htbp]
    \centering
    \includegraphics[width=\textwidth]{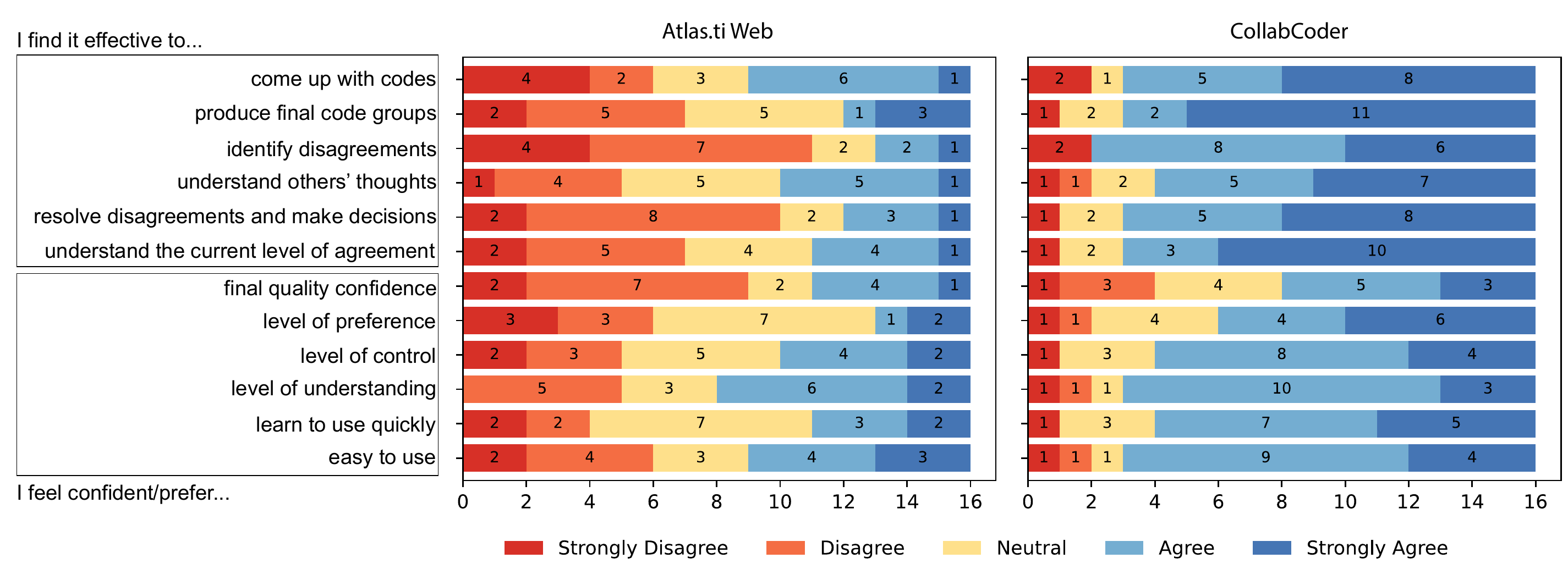}
    \caption{Post-study Questionnaires Responses from Our Participants on Different Dimensions on A 5-point Likert Scale, where 1 denotes "Strongly Disagree", 5 denotes "Strongly Agree". \added{The numerical values displayed on the stacked bar chart represent the count of participants who assigned each respective score.}}
    \label{fig:results}
\end{figure*}

\subsubsection{Key Findings (KF) on collaboration behaviors \added{with \diffcoder supports}}

\paragraph{\textbf{KF6: \added{An analysis of three intriguing group dynamics manifested in two conditions} \deleted{How \diffcoder workflow works under three different group dynamics}}}
\label{sec:results:KF6}

\added{In addition to the key findings on feature utilization,} we observed three intriguing collaboration group dynamics, including \textit{"follower-leader"} (P1$\times$P2, P5$\times$P6), \textit{"amicable cooperation"} (P3$\times$P4, P7$\times$P8, P9$\times$P10, P13$\times$P14, P15$\times$P16) and \textit{"swift but less cautious"} (P11$\times$P12). \added{The original observation notes are listed in Appendix Table \ref{tab:observation_notes_1} and \ref{tab:observation_notes_2}.}

The \textit{"follower-leader"} pattern typically occurred when one coder was a novice, while the other had more expertise. Often, the inexperienced coder contributed fewer ideas or only offered support during the coding process: when using Atlas.ti Web, those "lead" coders tended to take on more coding tasks than the others since their coding tasks could not be precisely quantified. Even though both of them were told to code all the data, it would end up in a situation where one coder primarily handled the work while the other merely followed with minimal input. This pattern could also appear if the coders worked at different paces (P1$\times$P2). As a result, the more efficient coders expressed more ideas. In contrast, \diffcoder ensures equitable participation by assigning the same coding workload to each participant and offering detailed documentation of the decision-making process via its independent coding interface. This approach guarantees that coders, even if they seldom voice their opinions directly, can still use the explicit documented information to communicate their ideas indirectly and be assessed in tandem with their collaborators. Furthermore, the suggestions generated by GPT are derived from both codes and raw data, producing a similar effect.

For \textit{"amicable cooperation"}, the coders respected each other's opinions \added{while employing \diffcoder as a collaborative tool to finalize their coding decisions.} When they make a decision, they firstly identify the common keywords between their codes, and then check the suggestions with similar keywords to decide whether to use suggestions or propose their own final code decision. Often, they took turns to apply the final code. 
For example, for the first data unit, one coder might say, \textit{"hey, mine seems better, let's use mine as the final decision,"} and for the second one, the coder might say, \textit{"hey, I like yours, we should choose yours [as the final decision]"} (P3$\times$P4). In some cases, such as P13$\times$P14, both coders generally reach a consensus, displaying no strong dominance and showing respect for each other's opinions, sometimes it is challenging to finalize a terminology for the code decision. Under this kind of condition, the coders used an LLMs agent as a mediator to find a more suitable expression that takes into account both viewpoints.
\added{Although most groups maintained similar "amicable cooperation" dynamics in Atlas.ti Web sessions, some found it challenging to adhere to their established patterns. This difficulty is attributed to the fact that such patterns are more resource-intensive. Take the P7$\times$P8 scenario as an example: in this case, the participants encountered time management challenges, as each coding session was initially scheduled to conclude within half an hour. Participants were afforded some flexibility, allowing sessions to extend slightly beyond the initially planned duration to ensure the completion of their tasks. In the \diffcoder condition, they engaged in extensive and respectful discussions, which consequently reduced the time available for the Atlas.ti Web session. Consequently, they had to expedite the process in Atlas.ti Web. This rush resulted in a situation where only one coder assumed the responsibility of merging the codes and rapidly grouping them into thematic clusters. 
For this coder, to access deeper insights behind these codes, additional operations like asking why another coder has this code, and clicking more to understand which sentence it means were often not feasible within the time constraints. This absence of operations forced coders to merge data relying solely on codes, without the advantage of additional contextual insights.
Consequently, this approach often leads to a "follower-leader" or "leader-takes-all" dynamic.
While this simplifies the process for participants, it potentially compromises the quality of the discussion. This is also evidenced by our quantitative data in Table \ref{tab:result}.}

The \textit{"swift but less cautious"} collaboration was a less desirable pattern we noticed: For P11$\times$P12, during the merging process, they would heavily rely on GPT-generated decisions in order to finish the task quickly. This scenario highlights the concerns regarding excessive reliance on GPT and insufficient deep thinking, which can negatively impact the final quality even when GPT is used as a mediator after the codes have been produced, as defined as our initial objective. Under this pattern, the pair sadly used GPT for "another round of coding" rather than as a neutral third-party decision advice provider. \added{In the case of this particular pair working with Atlas.ti Web, a distinct pattern emerged: P11 exhibited a notably faster pace, while P12 worked more slowly. As a result, the collaboration between the participants evolved into a "follower-leader" dynamic. In this structure, the quicker participant, P11, appeared to steer the overall process, occasionally soliciting inputs from P12.}

\subsection{\added{RQ2. How does \diffcoder compare to currently available tools like Atlas.ti Web?}}

\subsubsection{Post-study questionnaire}

We gathered the subjective preferences from our participants. To do so, we gave them 12 statements like \textit{"I find it effective to..."} and \textit{"I feel confident/prefer..."} pertaining to the effectiveness and self-perception. We then asked them to rate their agreement with each sentence on a 5-point Likert scale for each platform. The details of the 12 statements are shown in Figure~\ref{fig:results}.

Overall, pairwise t-tests showed that participants rated \diffcoder significantly (all $p<.05$) better than Atlas.ti Web for effectiveness in 1) \textit{coming up with codes}, 2) \textit{producing final code groups}, 3) \textit{identifying disagreements}, 4) \textit{resolving disagreements and making decisions}, 5) \textit{understanding the current level of agreement}, and 6) \textit{understanding others' thoughts}. The results also indicated that participants believed \diffcoder \added{($M=4$)} could be learned for use quickly compared to Atlas.ti Web \added{($M=3.1, t(15) = -3.05, p < .01$)}.
For other dimensions, the \textit{confidence in the final quality}, \textit{perceived level of preference}, \textit{level of control}, \textit{level of understanding}, and \textit{ease of use}, while our results show a general trend where \diffcoder achieves higher scores, we found no significant differences. 
Additionally, we observed that one expert user (P6) exhibited a highly negative attitude towards implementing AI in qualitative coding, as he selected "strongly disagree" for nearly all the assessment criteria. \added{We will discuss his qualitative feedback in Section \ref{sec:disc:helper}. }

\subsubsection{Log data analysis}
A two-tailed pairwise t-test on \textit{Discussion Time} revealed a significant difference ($t(15) = -3.22, p = .017$) between \diffcoder $(M\approx24mins, SD\approx7mins)$ and Atlas.ti Web $(M\approx11mins, SD\approx5.5mins)$. \textbf{Discussions under the \diffcoder condition were significantly longer than those in the Atlas.ti Web condition.}
When examining the IRR, it was found that the IRRs in the Atlas.ti Web condition were overall significantly ($t(7)=-6.69, p<.001$) lower $(M=0.06, SD =0.40)$, compared to the \diffcoder condition ($M \approx 1$).
In the latter, participants thoroughly examined all codes, resolved conflicts, merged similar codes, and reached a final decision for each data unit.
Conversely, Atlas.ti Web posed challenges in comparing individual data units side-by-side, leading to minimal code discussions overall (averaging 4.5 codes discussed) compared to the \diffcoder option (averaging 15 codes discussed). Consequently, we surmise that concealed disagreements within Atlas.ti Web might require additional discussion rounds to attain a higher agreement level. Further evidence is needed to validate this assumption.

\begin{table*}[!t]
\caption{Overview of the final coding results. "Collab." denotes \diffcoder, "Atlas." denotes Atlas.ti Web, "Total Codes" denotes the total number of codes generated while "Discussed Codes" denotes the total number of codes that were discussed by the coders during the discussion phase. "Bus." denotes the "Business" dataset while "His." denotes the "History" dataset. \added{"Suggestions Acceptance" column denotes the proportion of usage of GPT-generated codes (GPT), the selection from the relevant codes in code history suggested by GPT (Rele.), and users' self-proposed codes (Self.) to the total number of open codes in Phase 1. 
"GPT-based Code Decisions" column reflects the proportion of code decisions in Phase 2 that originated from suggestions made by the GPT mediator.}}
\label{tab:result}
\scalebox{0.66}{
\begin{tabular}{|cccc|cc|cc|cc|cc|cc|cc|c|c|}
\hline
\multicolumn{1}{|c|}{\multirow{2}{*}{\textbf{Pairs}}} & \multicolumn{1}{c|}{\multirow{2}{*}{\textbf{\begin{tabular}[c]{@{}c@{}}Self-reported\\ QA expertise\end{tabular}}}} & \multicolumn{1}{c|}{\multirow{2}{*}{\textbf{Conditions}}} & \multirow{2}{*}{\textbf{\begin{tabular}[c]{@{}c@{}}Collaboration\\ Observation\end{tabular}}} & \multicolumn{2}{c|}{\textbf{\begin{tabular}[c]{@{}c@{}}Total Codes\end{tabular}}} & \multicolumn{2}{c|}{\textbf{\begin{tabular}[c]{@{}c@{}}Discussed\\ Codes (No.)\end{tabular}}} & \multicolumn{2}{c|}{\textbf{\begin{tabular}[c]{@{}c@{}}IRR (-1 to 1)\end{tabular}}} & \multicolumn{2}{c|}{\textbf{\begin{tabular}[c]{@{}c@{}}Code \\Groups (No.)\end{tabular}}} & \multicolumn{2}{c|}{\textbf{\begin{tabular}[c]{@{}c@{}}Discussion \\ Time (mins:secs)\end{tabular}}} & \multicolumn{3}{c|}{\textbf{\begin{tabular}[c]{@{}c@{}}\small{\added{Suggestions Acceptance}} \\[-0.4em] \small{\added{in Phase 1} (\%)}\end{tabular}}} & \multirow{2}{*}{\textbf{\begin{tabular}[c]{@{}c@{}}\\[-1.5em]\small{\added{GPT-Based}}\\[-0.4em]\small{\added{Code}}\\[-0.4em] \small{\added{Decisions (\%)}}\end{tabular}}} \\ \cline{5-17}
\multicolumn{1}{|c|}{} & \multicolumn{1}{c|}{} & \multicolumn{1}{c|}{} &  & \multicolumn{1}{c|}{\textbf{Collab.}} & \textbf{Atlas.} & \multicolumn{1}{c|}{\textbf{Collab.}} & \textbf{Atlas.} & \multicolumn{1}{c|}{\textbf{Collab.\textsuperscript{b}}} & \textbf{Atlas.} & \multicolumn{1}{c|}{\textbf{Collab.}} & \textbf{Atlas.} & \multicolumn{1}{c|}{\textbf{Collab.}} & \textbf{Atlas.} & \multicolumn{1}{c|}{\textbf{GPT}} & \textbf{Rele.} & \textbf{Self.} &  \\ \hline
\multicolumn{1}{|c|}{P1} & \multicolumn{1}{c|}{Beginner} & \multicolumn{1}{c|}{\multirow{2}{*}{\begin{tabular}[c]{@{}c@{}}Atlas. (Bus.), \\ Collab.His.)\end{tabular}}} & \multirow{2}{*}{\begin{tabular}[c]{@{}c@{}}Follower-\\ Leader\end{tabular}} & \multicolumn{1}{c|}{\multirow{2}{*}{15}} & \multirow{2}{*}{24} & \multicolumn{1}{c|}{\multirow{2}{*}{15}} & \multirow{2}{*}{6} & \multicolumn{1}{c|}{\multirow{2}{*}{NA}} & \multirow{2}{*}{-0.07} & \multicolumn{1}{c|}{\multirow{2}{*}{6}} & \multirow{2}{*}{3} & \multicolumn{1}{c|}{\multirow{2}{*}{19:41}} & \multirow{2}{*}{07:39} & \multicolumn{1}{c|}{100} & 0 & 0 & \multirow{2}{*}{5} \\ \cline{15-17}
\multicolumn{1}{|c|}{P2} & \multicolumn{1}{c|}{No Experience} & \multicolumn{1}{c|}{} &  & \multicolumn{1}{c|}{} &  & \multicolumn{1}{c|}{} &  & \multicolumn{1}{c|}{} &  & \multicolumn{1}{c|}{} &  & \multicolumn{1}{c|}{} &  & \multicolumn{1}{c|}{70} & 5 & 25 &  \\ \hline
\multicolumn{1}{|c|}{P3} & \multicolumn{1}{c|}{Expert} & \multicolumn{1}{c|}{\multirow{2}{*}{\begin{tabular}[c]{@{}c@{}}Collab.(Bus.), \\ Atlas. (His.)\end{tabular}}} & \multirow{2}{*}{\begin{tabular}[c]{@{}c@{}}Amicable \\ Cooperation\end{tabular}} & \multicolumn{1}{c|}{\multirow{2}{*}{15}} & \multirow{2}{*}{10} & \multicolumn{1}{c|}{\multirow{2}{*}{15}} & \multirow{2}{*}{10} & \multicolumn{1}{c|}{\multirow{2}{*}{NA}} & \multirow{2}{*}{1} & \multicolumn{1}{c|}{\multirow{2}{*}{5}} & \multirow{2}{*}{4} & \multicolumn{1}{c|}{\multirow{2}{*}{35:24}} & \multirow{2}{*}{21:32} & \multicolumn{1}{c|}{90} & 0 & 10 & \multirow{2}{*}{40} \\ \cline{15-17}
\multicolumn{1}{|c|}{P4} & \multicolumn{1}{c|}{No Experience} & \multicolumn{1}{c|}{} &  & \multicolumn{1}{c|}{} &  & \multicolumn{1}{c|}{} &  & \multicolumn{1}{c|}{} &  & \multicolumn{1}{c|}{} &  & \multicolumn{1}{c|}{} &  & \multicolumn{1}{c|}{90} & 0 & 10 &  \\ \hline
\multicolumn{1}{|c|}{P5} & \multicolumn{1}{c|}{No Experience} & \multicolumn{1}{c|}{\multirow{2}{*}{\begin{tabular}[c]{@{}c@{}}Atlas. (His.), \\ Collab.(Bus.)\end{tabular}}} & \multirow{2}{*}{\begin{tabular}[c]{@{}c@{}}Follower-\\ Leader\end{tabular}} & \multicolumn{1}{c|}{\multirow{2}{*}{15}} & \multirow{2}{*}{11} & \multicolumn{1}{c|}{\multirow{2}{*}{15}} & \multirow{2}{*}{2} & \multicolumn{1}{c|}{\multirow{2}{*}{NA}} & \multirow{2}{*}{-0.02} & \multicolumn{1}{c|}{\multirow{2}{*}{5}} & \multirow{2}{*}{2} & \multicolumn{1}{c|}{\multirow{2}{*}{17:55}} & \multirow{2}{*}{06:16} & \multicolumn{1}{c|}{73} & 7 & 20 & \multirow{2}{*}{100} \\ \cline{15-17}
\multicolumn{1}{|c|}{P6} & \multicolumn{1}{c|}{Expert} & \multicolumn{1}{c|}{} &  & \multicolumn{1}{c|}{} &  & \multicolumn{1}{c|}{} &  & \multicolumn{1}{c|}{} &  & \multicolumn{1}{c|}{} &  & \multicolumn{1}{c|}{} &  & \multicolumn{1}{c|}{100} & 0 & 0 &  \\ \hline
\multicolumn{1}{|c|}{P7} & \multicolumn{1}{c|}{No Experience} & \multicolumn{1}{c|}{\multirow{2}{*}{\begin{tabular}[c]{@{}c@{}}Collab.(His.)\\ Atlas. (Bus.)\end{tabular}}} & \multirow{2}{*}{\begin{tabular}[c]{@{}c@{}}Amicable \\ Cooperation\end{tabular}} & \multicolumn{1}{c|}{\multirow{2}{*}{15}} & \multirow{2}{*}{22} & \multicolumn{1}{c|}{\multirow{2}{*}{15}} & \multirow{2}{*}{2} & \multicolumn{1}{c|}{\multirow{2}{*}{NA}} & \multirow{2}{*}{-0.33} & \multicolumn{1}{c|}{\multirow{2}{*}{7}} & \multirow{2}{*}{6} & \multicolumn{1}{c|}{\multirow{2}{*}{29:08}} & \multirow{2}{*}{\begin{tabular}[c]{@{}c@{}}No \\ discussion\textsuperscript{a}\end{tabular}} & \multicolumn{1}{c|}{7} & 0 & 93 & \multirow{2}{*}{80} \\ \cline{15-17}
\multicolumn{1}{|c|}{P8} & \multicolumn{1}{c|}{No Experience} & \multicolumn{1}{c|}{} &  & \multicolumn{1}{c|}{} &  & \multicolumn{1}{c|}{} &  & \multicolumn{1}{c|}{} &  & \multicolumn{1}{c|}{} &  & \multicolumn{1}{c|}{} &  & \multicolumn{1}{c|}{13} & 7 & 80 &  \\ \hline
\multicolumn{1}{|c|}{P9} & \multicolumn{1}{c|}{Intermediate} & \multicolumn{1}{c|}{\multirow{2}{*}{\begin{tabular}[c]{@{}c@{}}Atlas. (Bus.), \\ Collab.(His.)\end{tabular}}} & \multirow{2}{*}{\begin{tabular}[c]{@{}c@{}}Amicable \\ Cooperation\end{tabular}} & \multicolumn{1}{c|}{\multirow{2}{*}{15}} & \multirow{2}{*}{17} & \multicolumn{1}{c|}{\multirow{2}{*}{15}} & \multirow{2}{*}{5} & \multicolumn{1}{c|}{\multirow{2}{*}{NA}} & \multirow{2}{*}{0.04} & \multicolumn{1}{c|}{\multirow{2}{*}{5}} & \multirow{2}{*}{2} & \multicolumn{1}{c|}{\multirow{2}{*}{15:11}} & \multirow{2}{*}{14:38} & \multicolumn{1}{c|}{73} & 13 & 13 & \multirow{2}{*}{80} \\ \cline{15-17}
\multicolumn{1}{|c|}{P10} & \multicolumn{1}{c|}{No Experience} & \multicolumn{1}{c|}{} &  & \multicolumn{1}{c|}{} &  & \multicolumn{1}{c|}{} &  & \multicolumn{1}{c|}{} &  & \multicolumn{1}{c|}{} &  & \multicolumn{1}{c|}{} &  & \multicolumn{1}{c|}{53} & 40 & 7 &  \\ \hline
\multicolumn{1}{|c|}{P11} & \multicolumn{1}{c|}{Intermediate} & \multicolumn{1}{c|}{\multirow{2}{*}{\begin{tabular}[c]{@{}c@{}}Collab.(Bus.), \\ Atlas. (His.)\end{tabular}}} & \multirow{2}{*}{\begin{tabular}[c]{@{}c@{}}Quick and\\ not careful\end{tabular}} & \multicolumn{1}{c|}{\multirow{2}{*}{15}} & \multirow{2}{*}{61} & \multicolumn{1}{c|}{\multirow{2}{*}{15}} & \multirow{2}{*}{2} & \multicolumn{1}{c|}{\multirow{2}{*}{NA}} & \multirow{2}{*}{-0.07} & \multicolumn{1}{c|}{\multirow{2}{*}{3}} & \multirow{2}{*}{3} & \multicolumn{1}{c|}{\multirow{2}{*}{19:23}} & \multirow{2}{*}{14:15} & \multicolumn{1}{c|}{100} & 0 & 0 & \multirow{2}{*}{100} \\ \cline{15-17}
\multicolumn{1}{|c|}{P12} & \multicolumn{1}{c|}{No experience} & \multicolumn{1}{c|}{} &  & \multicolumn{1}{c|}{} &  & \multicolumn{1}{c|}{} &  & \multicolumn{1}{c|}{} &  & \multicolumn{1}{c|}{} &  & \multicolumn{1}{c|}{} &  & \multicolumn{1}{c|}{100} & 0 & 0 &  \\ \hline
\multicolumn{1}{|c|}{P13} & \multicolumn{1}{c|}{Beginner} & \multicolumn{1}{c|}{\multirow{2}{*}{\begin{tabular}[c]{@{}c@{}}Atlas. (His.), \\ Collab.(Bus.)\end{tabular}}} & \multirow{2}{*}{\begin{tabular}[c]{@{}c@{}}Amicable \\ Cooperation\end{tabular}} & \multicolumn{1}{c|}{\multirow{2}{*}{15}} & \multirow{2}{*}{30} & \multicolumn{1}{c|}{\multirow{2}{*}{15}} & \multirow{2}{*}{5} & \multicolumn{1}{c|}{\multirow{2}{*}{NA}} & \multirow{2}{*}{-0.08} & \multicolumn{1}{c|}{\multirow{2}{*}{8}} & \multirow{2}{*}{2} & \multicolumn{1}{c|}{\multirow{2}{*}{29:19}} & \multirow{2}{*}{08:43} & \multicolumn{1}{c|}{87} & 7 & 7 & \multirow{2}{*}{100} \\ \cline{15-17}
\multicolumn{1}{|c|}{P14} & \multicolumn{1}{c|}{Intermediate} & \multicolumn{1}{c|}{} &  & \multicolumn{1}{c|}{} &  & \multicolumn{1}{c|}{} &  & \multicolumn{1}{c|}{} &  & \multicolumn{1}{c|}{} &  & \multicolumn{1}{c|}{} &  & \multicolumn{1}{c|}{93} & 0 & 7 &  \\ \hline
\multicolumn{1}{|c|}{P15} & \multicolumn{1}{c|}{Beginner} & \multicolumn{1}{c|}{\multirow{2}{*}{\begin{tabular}[c]{@{}c@{}}Collab.(His.)\\ Atlas. (Bus.)\end{tabular}}} & \multirow{2}{*}{\begin{tabular}[c]{@{}c@{}}Amicable \\ Cooperation\end{tabular}} & \multicolumn{1}{c|}{\multirow{2}{*}{15}} & \multirow{2}{*}{8} & \multicolumn{1}{c|}{\multirow{2}{*}{15}} & \multirow{2}{*}{4} & \multicolumn{1}{c|}{\multirow{2}{*}{NA}} & \multirow{2}{*}{0.04} & \multicolumn{1}{c|}{\multirow{2}{*}{4}} & \multirow{2}{*}{2} & \multicolumn{1}{c|}{\multirow{2}{*}{29:09}} & \multirow{2}{*}{08:52} & \multicolumn{1}{c|}{100} & 0 & 0 & \multirow{2}{*}{43} \\ \cline{15-17}
\multicolumn{1}{|c|}{P16} & \multicolumn{1}{c|}{No experience} & \multicolumn{1}{c|}{} &  & \multicolumn{1}{c|}{} &  & \multicolumn{1}{c|}{} &  & \multicolumn{1}{c|}{} &  & \multicolumn{1}{c|}{} &  & \multicolumn{1}{c|}{} &  & \multicolumn{1}{c|}{73} & 20 & 7 &  \\ \hline
\multicolumn{4}{|c|}{Mean} & \multicolumn{1}{c|}{15} & 22.88 & \multicolumn{1}{c|}{15} & 4.5 & \multicolumn{1}{c|}{NA} & 0.06 & \multicolumn{1}{c|}{5.38} & 3 & \multicolumn{1}{c|}{24:00} & 10:48 & \multicolumn{1}{c|}{76.46} & 6.15 & 17.4 & 68.5 \\ \hline
\multicolumn{4}{|c|}{SD} & \multicolumn{1}{c|}{0} & 17.19 & \multicolumn{1}{c|}{0} & 2.73 & \multicolumn{1}{c|}{NA} & 0.40 & \multicolumn{1}{c|}{1.60} & 1.41 & \multicolumn{1}{c|}{07:12} & 05:24 & \multicolumn{1}{c|}{29.43} & 10.74 & 28.11 & 35.3 \\ \hline
\end{tabular}
}

\vspace{1ex}
\raggedright
\footnotesize
\textsuperscript{a} P7 and P8 gave up discussion for the Atlas.ti session due to spending too much time in the \diffcoder session. \\
\textsuperscript{b} \added{Following the discussion session in \diffcoder, the original codes have been restructured and finalized as a single code decision, resulting in an IRR$\approx$1. Consequently, IRR calculations are not applicable (NA) for the \diffcoder conditions.} 

\end{table*}

\subsection{\added{RQ3. How can the design of CollabCoder be improved?}}

\label{sec:results:attitudes}

\added{While \diffcoder effectively facilitates collaboration in various aspects, as discussed in Section \ref{sec:results:rq1}, we observed divergent attitudes toward certain functions, such as labeling certainty, relevant code suggestions, and the use of individual codebooks.}

Most participants expressed concerns about the clarity, usefulness, and importance of the certainty function in \diffcoder. The self-reported nature, the potential of inconsistencies in reporting, and minimal usage among users suggest that the \textbf{certainty function may not be as helpful as intended}. For example, P12 found the certainty function "not really helpful", and P13 admitted forgetting about it due to the numerous other subtasks in the coding process. P3 also reported limited usage of the function, mainly assigning low certainty scores when not understanding the raw data. However, P14 recognized that the certainty function could be helpful in larger teams, as it might help flag quotes that require more discussion. 

The perceived usefulness of the relevant code function in \diffcoder depends on the dataset and users' preferences. Some participants found it \textbf{less relevant than the AI agent's summary function}, which they considered more accurate and relevant. \textit{"Maybe not that useful, but I think it depends on your dataset. Say whether they are many similar data points or whether they are different data points. So I think in terms of these cases they are all very different, have a lot of different contents. So it's not very relevant, but definitely, I think, in datasets which might be more relevant, could be useful."} (P2)

As for the individual codebook function, although users acknowledged its potential usefulness in tracking progress and handling large datasets, most users \textit{"did not pay much attention to it during this coding process"} (P2, P3, P4). 
P3 found it helpful for tracking progress but did not pay attention to it during this particular process. P4 acknowledged that the function could be useful in the long run, particularly when dealing with a large amount of data.

\added{While these features may not be as useful as initially anticipated, evidenced by low usage frequency or varying effectiveness across different datasets, further investigation is necessary to ascertain if the needs and challenges associated with these features truly exist or are merely perceived by us. This could significantly enhance user experiences with \diffcoder and inform the future design of AI-assisted CQA tools.}

\section{Discussion and Design Implications}

\subsection{Facilitating Rigorous, Lower-barrier CQA Process through Workflow Design Aligned with Theories}

\added{Practically, \diffcoder contributes by providing a one-stop, end-to-end workflow that ensures seamless data transitions between stages with minimal effort. This design is grounded in qualitative analysis theories such as Grounded Theory~\cite{flick2013sage} and Thematic Analysis~\cite{maguire2017doing}, as outlined in Section \ref{sec:background_theory}, facilitating a rigorous yet accessible approach to CQA practice. While spreadsheets are also capable of similar processes, they typically demand considerable effort and struggle to uphold a stringent process due to the intricacy and nuances involved. \diffcoder, in contrast, streamlines these tasks, rendering the team coordination process~\cite{entin1999adaptive, malone1994coordination} more practical and manageable. Our evaluation demonstrates the effectiveness of \diffcoder, empowering both experienced practitioners and novices to perform rigorous and comprehensive qualitative analysis. }

\added{Apart from practical benefits, our \diffcoder design \cite{CASH201884} can also enrich theoretical understanding in the CQA domain \cite{atlas}, which aids practitioners in grasping foundational theories, thereby bolstering the credibility of qualitative research \cite{collins2018international, atlas}. Over the years, CQA practices have remained inconsistent and vague, particularly regarding when and how multiple coders may be involved, the computation of IRR, the use of individual coding phases, and adherence to existing processes~\cite{bradley1993methodological, noble}. A common question could arise: \textit{"If deviating from strict processes does not significantly impact results, or the influence is hard to perceive (at least from others' perspective), why should substantial time be invested in maintaining them, especially under time constraints?"} Current software like Atlas.ti, MaxQDA often neglects this critical aspect in their system design, focusing instead on basic functionalities like data maintenance and code addition, which, are not the most challenging parts of the process for practitioners. 
Ultimately, \diffcoder enables practitioners to conduct a CQA process that is both transparent and standardized within the community~\cite{noble, moravcsik_2014}. Looking forward, we foresee a future where coders, in documenting their methodologies, will readily reference their use of such specifically designed workflows or systems for CQA analysis.}

\added{With this in mind, our objective is not to position any single method as the definitive standard in this field. Although \diffcoder is specifically designed for one type of coding — consensus coding within inductive coding — we do not exclusively advocate for either consensus or split coding. Instead, we emphasize that coders should choose a method that aligns best with their data and requirements~\cite{teherani2015choosing, atlas, gradcoach_2023_qualitative, collins2018international}. Therefore, the design of such tools should aim to accommodate various types of qualitative analysis methods. For instance, split coding might necessitate distributing data among team members in a manner that differs from the uniform distribution required by consensus coding.}

\subsection{LLMs as ``Suggestion Provider'' in Open Coding: Helper, not Replacement.}
\label{sec:disc:llm_suggestion_provider}

\subsubsection{Utilizing LLMs to Reduce Cognitive Burden}
Independent open coding is a highly cognitively demanding task, as it requires understanding the text, identifying the main idea, creating a summary based on research questions, and formulating a suitable phrase to convey the summary~\cite{lazar2017research, corbin2009basic}. Additionally, there is the need to refer to and reuse previously created codes. In this context, GPT's text comprehension and generation capabilities can assist in this mentally challenging process by serving as a suggestion provider.

\subsubsection{Improving LLMs' Suggestions Quality}
However, a key consideration according to \hyperref[sec:results:KF2]{KF2} is how GPT can provide better quality suggestions that align with the needs of users. For \diffcoder, we only provided essential prompts such as "summary" and "relevant codes". However, a crucial aspect of qualitative coding is that coders should always consider their research questions while coding and work towards a specific direction. For instance, are they analyzing the main sentiment of the raw data or the primary opinion regarding something? This factor can significantly impact the coding approach (e.g., descriptive or in-vivo coding~\cite{saldana2021coding}) and what should be coded (e.g., sentiment or opinions). Therefore, the system should support mechanisms for users to inform GPT of the user's intent or direction. One possible solution is to include the research question or intended direction in the prompt sent to GPT alongside the data to be coded. Alternatively, users could configure a customized prompt for guidance, directing GPT's behavior through the interface~\cite{ippolito2022creative}. This adaptability accommodates individual preferences and improves the overall user experience.

\added{Looking ahead, as the underlying LLM evolves, we envision that an approach for future LLM assistance in CollabCoder involves: 1) creating a comprehensive library of both pre-set and real-time updated prompts, designed to assist in suggesting codes across diverse fields like psychology and HCI; 2) implementing a feature that allows coders to input custom prompts when the default prompts are not suitable.}

\subsubsection{LLMs should Remain a Helper}
\label{sec:disc:helper}
Another key consideration is how GPT can stay a reliable suggestion provider without taking over from the coder~\cite{jiang2021supporting, marathe2018}. Our study demonstrated that both novices and experts valued GPT's assistance, as participants used GPT's suggestions either as code or as a basis to create codes 76.67\% of the time on average.

\added{However, one expert user (P6) held a negative attitude towards employing LLMs in open coding, assigning the lowest score to nearly all measures (see Figure \ref{fig:results}). This user expressed concerns about the role of AI in this context, suggesting that qualitative researchers might feel forced to use AI-generated codes, which could introduce potential biases. Picking up the nuances from the text is considered \textit{"fun"} for qualitative researchers (P6), and suggestions should not give the impression that \textit{"the code is done for them and they just have to apply it"} (P6) or lead them to \textit{"doubt their own ideas"} (P5).}

On the other side, it is important not to overlook the risk of over-reliance on GPT. While we want GPT to provide assistance, we do not intend for it to fully replace humans in the process, as noted in \hyperref[sec:design_goal_5]{DG5}. Our observations revealed that although participants claimed they would read the raw data first and then check GPT's suggestions, some beginners tended to rely on GPT for forming their suggestions, and experts would unconsciously accept GPT's suggestions if unsure about the meaning of the raw data, in order to save time.  
Therefore, preserving the enjoyment of qualitative research and designing for appropriate reliance~\cite{lee2004trust} to avoid misuse~\cite{dzindolet2003role} or over-trust can be a complex challenge~\cite{xiao2023supporting}. To this end, mixed-initiative systems~\cite{allen1999mixed, horvitz1999principles} like \diffcoder can be designed to allow for different levels of automation. For example, GPT-generated suggestions could be provided only for especially difficult cases upon request, rather than being easily accessible for every unit, even when including a pre-defined time delay.

\subsection{LLMs as ``Mediator'' and ``Facilitator'' in Coding Discussion}

Among the three critical CQA phases we pinpointed, aside from the open coding phase, the subsequent two stages — Phase 2 (merge and discussion) and Phase 3 (development of a codebook) — require a shared workspace for coders to converse. We took note of the role LLMs undertook during these discussions.

\subsubsection{LLMs as a ``Mediator'' in Group Decision-Making.}
The challenge of dynamically reaching consensus — a decision that encapsulates the perspectives of all group members — has garnered attention in the HCI field~\cite{emamgholizadeh2022supporting, perez_dynamic_2018, jiang2021supporting}. \added{Jiang et al. \cite{jiang2021supporting} extensively explore collaborative dynamics in their research for qualitative analysis. They highlighted decision-making modes in consensus-building may vary under different power dynamics~\cite{iisc_2018_power_dynamics} in CQA context. In some cases, the primary author or a senior member of a project may assume the decision-making role. According to our \hyperref[sec:results:KF6]{KF6}, we also found interesting group dynamics, identifying patterns like "amicable cooperation", "follower-leader", and "swift but less cautious" modes. Our design positions GPT as a mediator or a group recommendation system ~\cite{emamgholizadeh2022supporting}, particularly useful when consensus is hard to reach. In this role, GPT acts as an impartial facilitator, aiding in harmonizing labor distribution and opinion expression. It guides groups towards decisions that are not only cost-effective but also equitable, justified, and sound~\cite{chen2013human}. This is a functionality that can hardly be achieved by using tools like Atlas.ti Web.
In fact, these group dynamics can also be explored through various lenses, such as the Thomas-Kilmann conflict modes~\cite{thomas2008thomas}, which emphasize the importance of balancing assertiveness and cooperativeness in a team. Delving into these theories can significantly aid in the design of more effective team collaboration tools. }

Nonetheless, \diffcoder's present design in Phase 2, which employs LLMs as a recommendation system for coding decisions, represents merely an initial step. \added{While the \diffcoder cannot fundamentally alter collaborative power dynamics, it ensures that coding is a collaborative effort, emphasizing substantive discussions between two coders to avoid superficial collaboration.} Looking ahead, there are numerous paths we could and should pursue. For example, as humans should be the ultimate decision-makers, with GPT serving merely as a fair mediator between coders, group decision recommendations ought to be made available only upon explicit request. Alternatively, once a coder puts forth a final decision, GPT could then refine the wording or formulate some conclusive description to facilitate future reflection on the code decisions~\cite{barry1999using}.

\subsubsection{LLMs as ``Facilitator'' in Streamlining Primary Code Grouping}
As per \hyperref[sec:results:KF5]{KF5}, our participants offered insightful feedback about using GPT to generate primary code groups. They found the top-down approach, where GPT first generates primary groups and users subsequently refine and revise them, more efficient and less cognitively demanding compared to the traditional bottom-up method. In the traditional method, users must begin by examining all primary codes, merging them, and then manually grouping them into categories, which can be mentally taxing. \added{Differently, \diffcoder is designed to initially formulate primary or coarse ideas about how to group codes. Similar to many types of recommendation systems, the suggestions provided by \diffcoder are intended to complement the coders' initial thoughts on code grouping. When coders review these GPT-suggested code groups, it enables them to reflect upon and compare their own ideas with the given suggestions. This process enriches the final code groups by efficiently incorporating a wider range of perspectives, extending beyond the insights of just the two coders. This ensures a more comprehensive and multifaceted categorization.} 
Moreover, researchers can more effectively and easily manage large volumes of data and potentially enhance the quality of their analysis.

However, it is crucial to exercise caution when applying this method. We observed that when time constraints exist, coders may skip discussions, with only one of two coders combining and categorizing the codes into code groups (P7$\times$P8). Additionally, P14 mentioned that GPT appears to dominate the code grouping process, resulting in a single approach to grouping. For instance, while the participants might create code groups based on sentiment analysis during their own coding process, they could be tempted to focus on content analysis under GPT's guidance.

Similarly, to overcome these challenges of \diffcoder, we envision a system where coders would create their own groupings first and only request LLMs' suggestions afterward. Alternatively, LLMs' assistance could be limited to situations where the data volume is substantial. Another approach could be prompting LLMs to generate code groups based on the research questions rather than solely on the (superficial) codes. This would ensure a more contextually relevant and research-driven code grouping process.

\section{Limitations and Future Work}
This work has limitations. \textbf{Firstly, it's important to note that the current version of \diffcoder operates under certain assumptions, deeming coding tasks as "ideal" — comprising semantically independent units, a two-person coding team, and data units with singular semantics.} However, our expert interviews revealed a more complex reality. One primary source of disagreement arises when different users assign multiple codes to the same data unit, often sparking discussions during collaborative coding. Future research should aim to address this point. 

Secondly, we only used pre-defined unit data and did not consider splitting complex data into units (e.g., interview data). Future work could explore utilizing GPT to support the segmentation of interview data into semantic units and automating the import process.

Lastly, we did not investigate the specific process by which users select and edit a GPT suggestion. Future research could delve deeper into how users incorporate these suggestions to generate a final idea. \added{The optimal time for balancing user autonomy and appropriate reliance should also be explored.} Moreover, for a tool that could be used by the same coder on multiple large datasets, it would also be beneficial to have GPT generate suggestions based on users' coding patterns rather than directly providing suggestions.

\section{Conclusion}
This paper introduces \diffcoder, a system that integrates the key stages of the CQA process into a one-stop workflow, aiming to lower the bar for adhering to a strict CQA procedure. Our evaluation with 16 participants indicated a preference for \diffcoder over existing platforms like Atlas.ti Web due to its user-friendly design and AI assistance tailored for collaboration. We also demonstrated the system's capability to streamline and facilitate discussions, guide consensus-building, and create codebooks. By examining both human-AI and human-human interactions within the context of qualitative analysis, we have uncovered key challenges and insights that can guide future design and research.

\bibliographystyle{ACM-Reference-Format}
\bibliography{reference}


\newpage
\appendix

\newcolumntype{T}{>{\centering\arraybackslash}m{0.1\linewidth}}
\newcolumntype{Y}{>{\centering\arraybackslash}m{0.5\linewidth}}
\newcolumntype{Z}{>{\centering\arraybackslash}m{0.3\linewidth}}

\begin{table*}[!htbp]
\centering
\caption{Different CQA Software. Note: This list is based on public online resources and not exhaustive.}
\label{tab:appendix:CQA_software}
\scalebox{0.8}{\begin{tabular}{|l|l|l|l|l|l|}
\hline
\textbf{Application} & \textbf{Atlas.ti Desktop} & \textbf{Atlas.ti Web} & \textbf{NVivo Desktop} & \textbf{Google docs} & \textbf{MaxQDA} \\ \hline
\textbf{\begin{tabular}[c]{@{}l@{}}Collaboration\\ ways\end{tabular}} & \begin{tabular}[c]{@{}l@{}}Coding separately and \\ then export the project \\ bundles to other coders\end{tabular} & \begin{tabular}[c]{@{}l@{}}Coding on the \\ same web page\end{tabular} & \begin{tabular}[c]{@{}l@{}}Coding separately and \\ then export the project \\ bundles to other coders\end{tabular} & \begin{tabular}[c]{@{}l@{}}Collaborative \\simultaneously\end{tabular} & \begin{tabular}[c]{@{}l@{}}Provide master project \\ that includes documents \\ and primary codes, and \\ then send copies to others, \\ allowing them to merge\end{tabular} \\ \hline
\textbf{Coding phase} & \begin{tabular}[c]{@{}l@{}}All Phases\end{tabular} & All phases & All Phases & All phases & All Phases \\ \hline
\textbf{Independence} & independent & not independent & Inpedendent & not independent & Inpedendent \\ \hline
\textbf{Synchrony} & Asynchronous & Synchronous & Asynchronous & Synchronous & Asynchronous \\ \hline
\textbf{Unit of analysis} & \begin{tabular}[c]{@{}l@{}}Select any text \end{tabular} & Select any text & \begin{tabular}[c]{@{}l@{}}Select any text, but \\ calculation of IRR can be \\ on character, sentence, \\ paragraph\end{tabular} & Select any text & Select any text \\ \hline
\textbf{IRR} & \begin{tabular}[c]{@{}l@{}}Agreement Percentage; \\ Holsti Index; \\ Krippendorff's \\ family of Alpha\end{tabular} & NA & \begin{tabular}[c]{@{}l@{}}Agreement Percentage; \\ Kappa coefficient\end{tabular} & NA & \begin{tabular}[c]{@{}l@{}}Agreement Percentage; \\ Kappa coefficient\end{tabular} \\ \hline
\textbf{\begin{tabular}[c]{@{}l@{}}Calculation \\ of IRR\end{tabular}} & \begin{tabular}[c]{@{}l@{}}Calculating after coding \\ system is stable and \\ all codes are defined\end{tabular} & \begin{tabular}[c]{@{}l@{}}Calculating \\manually\\ at any time\end{tabular} & \begin{tabular}[c]{@{}l@{}}Calculating after coding \\ system is stable and \\ all codes are defined \end{tabular} & NA & \begin{tabular}[c]{@{}l@{}}Calculating after coding \\ system is stable and \\ all codes are defined\end{tabular} \\ \hline
\textbf{\begin{tabular}[c]{@{}l@{}}Multi-valued \\ coding\end{tabular}} & \begin{tabular}[c]{@{}l@{}}support multiple  \\codes\end{tabular} & \begin{tabular}[c]{@{}l@{}}support multiple  \\codes\end{tabular} & \begin{tabular}[c]{@{}l@{}}support multiple \\ codes\end{tabular} & \begin{tabular}[c]{@{}l@{}}support multiple  \\codes\end{tabular} & \begin{tabular}[c]{@{}l@{}}support multiple \\ codes\end{tabular} \\ \hline
\textbf{\begin{tabular}[c]{@{}l@{}}Uncertainty/\\Disagreements\end{tabular}} & NA & NA & \begin{tabular}[c]{@{}l@{}}quickly identify areas of \\ agreement and disagreement \\ within the source data \\ using the green, yellow, \\ and blue indicators on the \\ scroll bar.\end{tabular} & NA & \begin{tabular}[c]{@{}l@{}}NA\end{tabular} \\ \hline
\end{tabular}}
\end{table*}

\begin{figure*}[!htbp]
    \centering
    \fbox{\includegraphics[width=\textwidth]{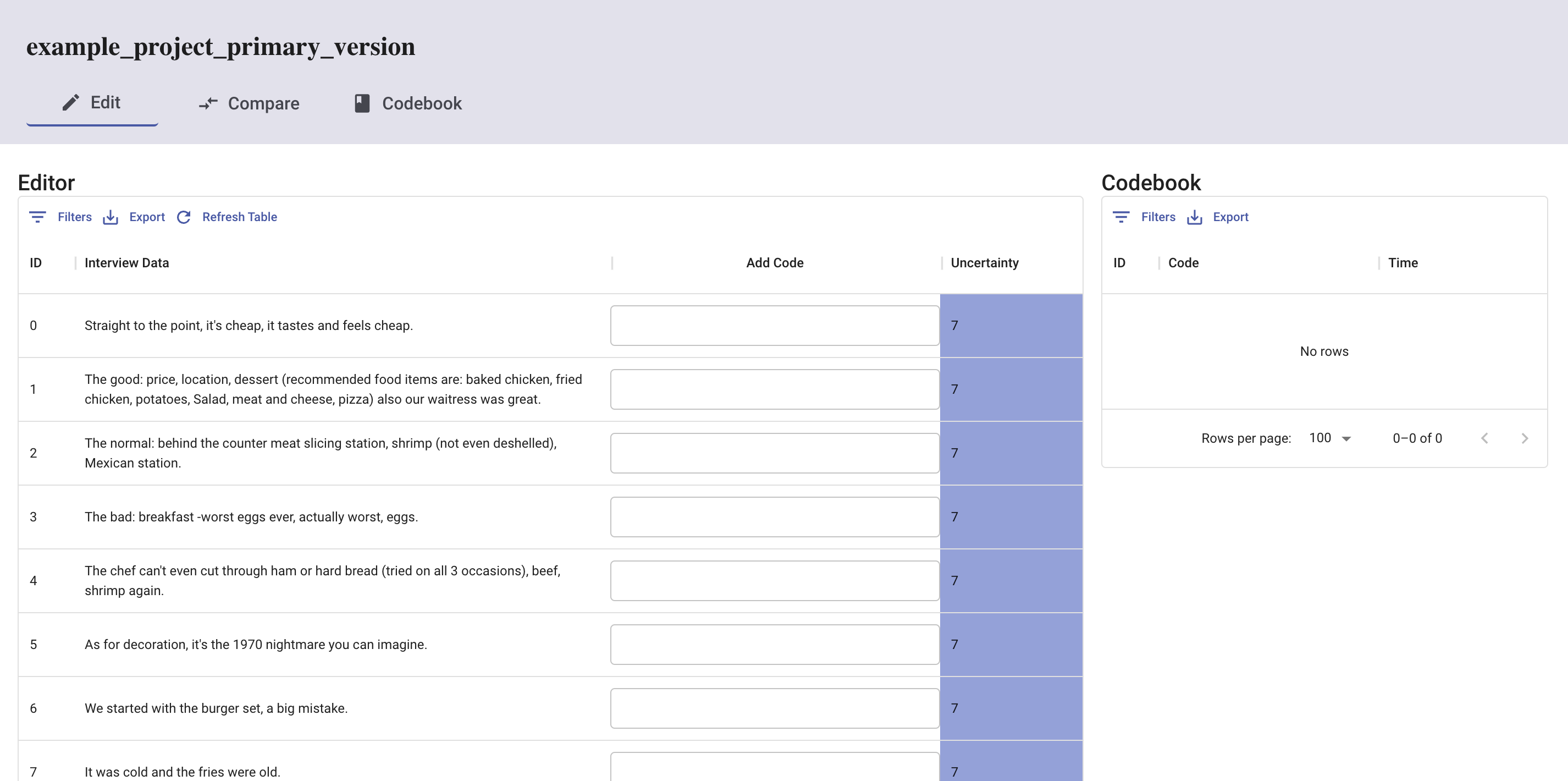}}
    \caption{Primary Prototype for Phase 1.}
    \label{fig:appendix:phase1}
\end{figure*}

\begin{figure*}[!htbp]
    \centering
    \fbox{\includegraphics[width=\textwidth]{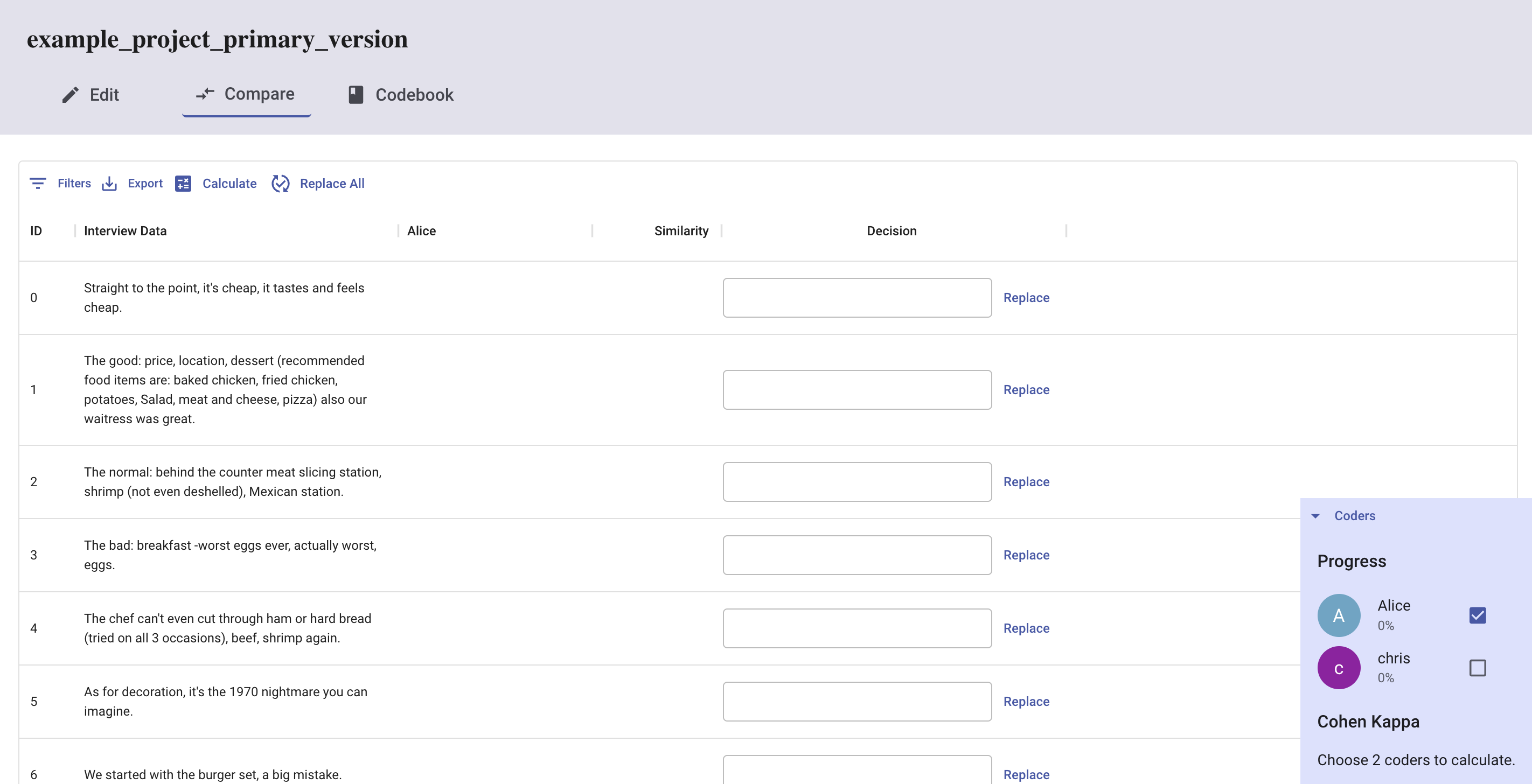}}
    \caption{Primary Prototype for Phase 2.}
    \label{fig:appendix:phase2}
\end{figure*}

\begin{figure*}[!htbp]
    \centering
    \fbox{\includegraphics[width=\textwidth]{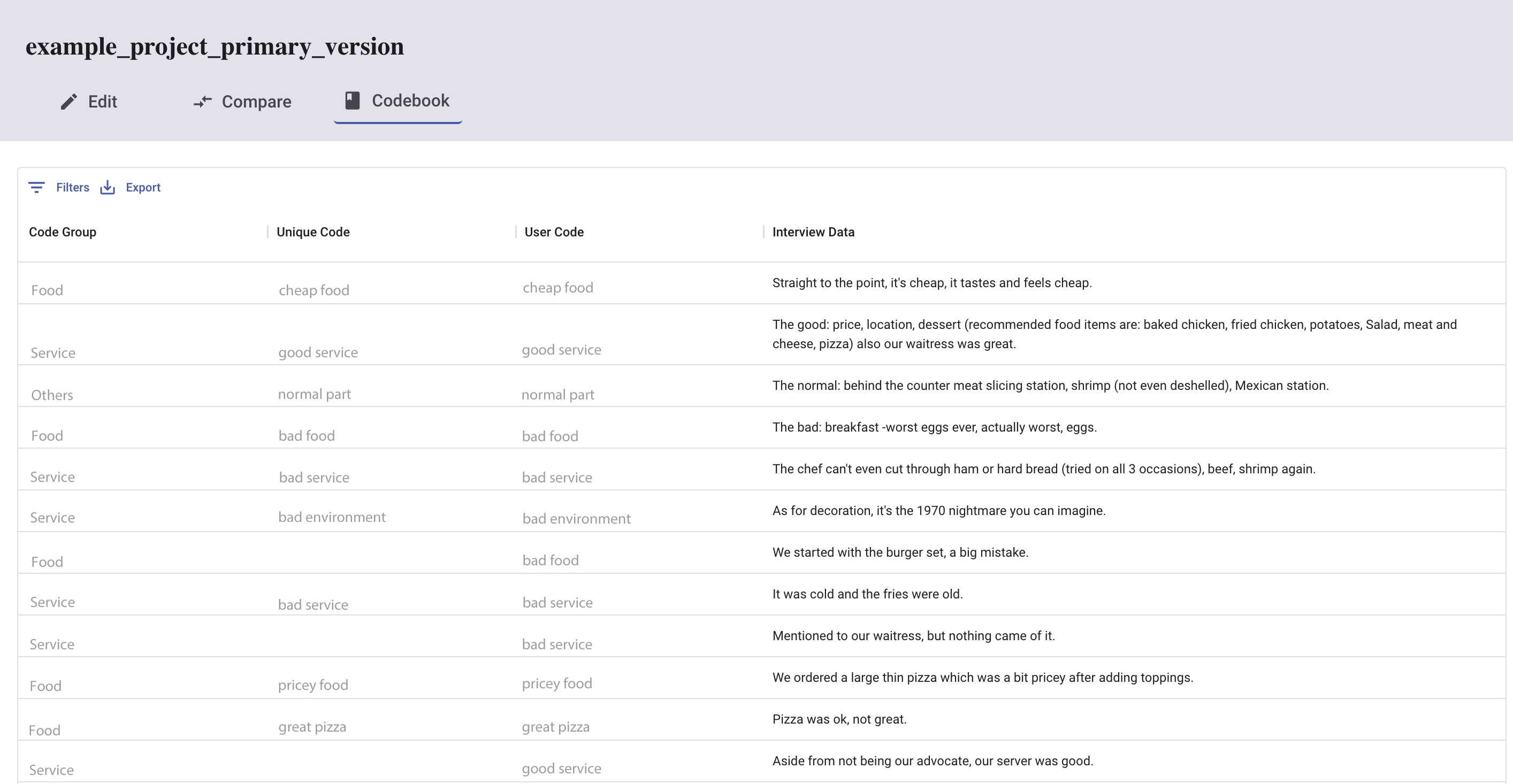}}
    \caption{Primary Prototype for Phase 3. \added{The gray-colored codes serve as an example to illustrate the differences between "Code Group", "Unique Code", and "User Code". The interfaces shown above, being preliminary mockups, were utilized to gather feedback from our primary interviewees in Step3, Section \ref{sec:DGs:method}, for the refinement of the final version of interfaces including Figure \ref{fig:interface:editing}, \ref{fig:interface:compare}, and \ref{fig:interface:codegroup}.}}
    \label{fig:appendix:phase3}
\end{figure*}

\begin{figure*}[!htbp]
    \centering
    \includegraphics[width=\textwidth]{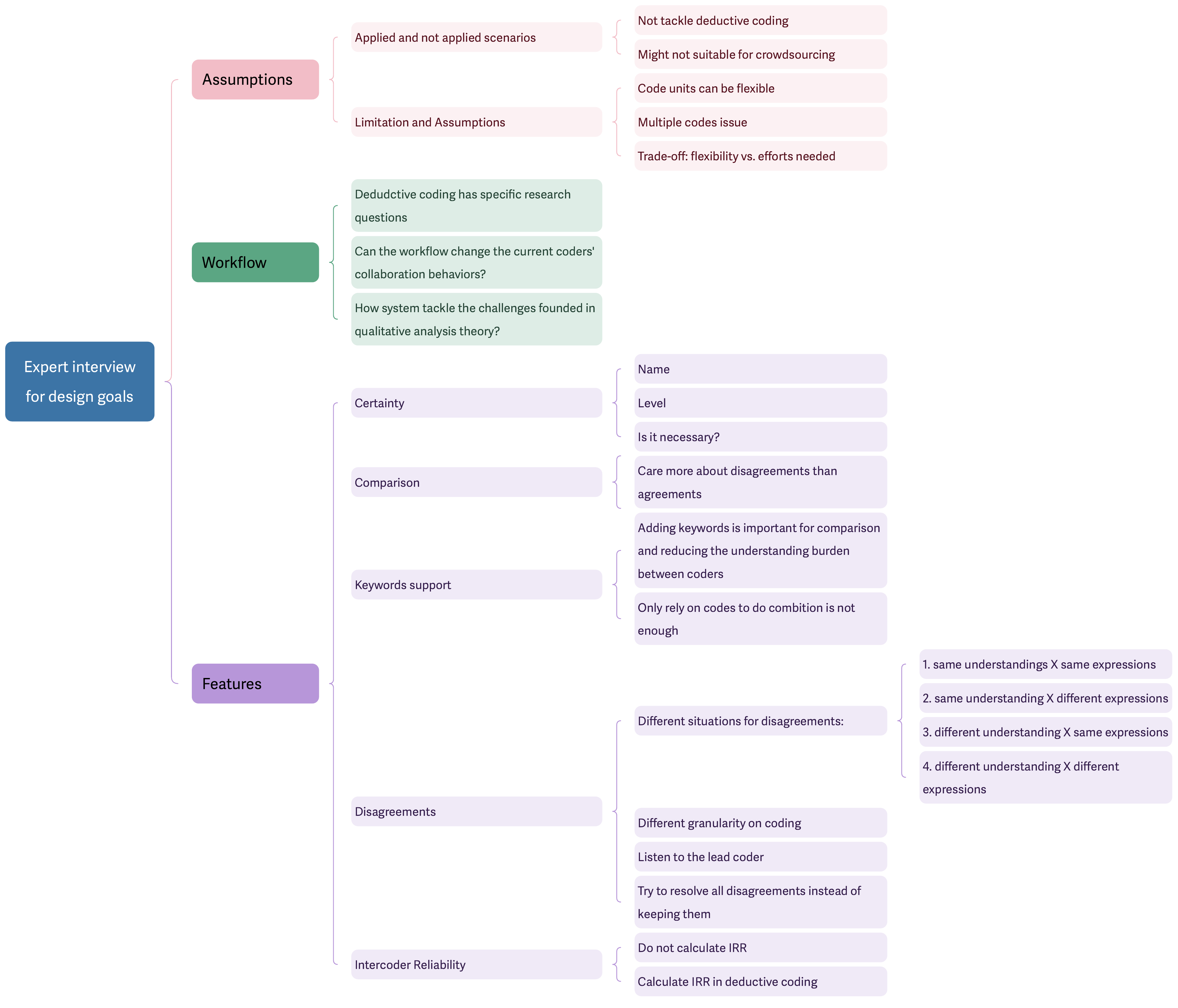}
    \caption{\added{Results of thematic analysis in Step3 (Section \ref{sec:DGs:method}) from expert interviews to derive design goals, with each node representing a coded element.}}
    \label{fig:appendix:expert_interview}
\end{figure*}

\begin{table*}[!htbp]
\caption{The prompts utilized in \diffcoder in Phase 1 when communicating with the ChatGPT API to produce code suggestions for text.}
\label{tab:system:prompt1}
\scalebox{0.8}{\begin{tabular}{|l|l|l|l|}
\hline
Phases & Features & Prompt Template & Example \\ \hline
\multirow{2}{*}{Phase 1} & \begin{tabular}[c]{@{}l@{}}Seek code \\ suggestions \\ for units\end{tabular} & \begin{tabular}[c]{@{}l@{}}$\bullet$ \textbf{system role}: You are a helpful qualitative \\ analysis assistant,  aiding researchers in \\ developing codes that can be utilized in \\ subsequent stages, including discussions \\ for creating codebooks and final coding \\ processes;\\ $\bullet$ \textbf{user input}: Please create three general \\ summaries for {[}text{]} (within six-word);\end{tabular} & \begin{tabular}[c]{@{}l@{}}{[}Text{]}:\\ "How A Business Works was an excellent book \\ to read as I began my first semester as a college \\ student. Although my goal is to major in Business, \\ I started my semester off with no idea of even the \\ basic guidelines a Business undergrad should know. \\ This book describes in detail every aspect dealing \\ with business relations, and I enjoyed reading it. \\ It felt great going to my additional business classes \\ prepared and knowledgeable on the subject they \\ were describing. Very well written, Professor \\ Haeberle! I recommend this book to anyone and \\ everyone who would like additional knowledge \\ pertaining to business matters."\\ \\ \textbf{Three general summaries for the above} {[}\textbf{Text}{]}:\\ 1. Book enlightened my initial business journey.\\ 2. Comprehensive guide for business undergraduates.\\ 3. Knowledge boost for new business students.\end{tabular} \\ \cline{2-4} 
 & \begin{tabular}[c]{@{}l@{}}Seek most \\ relevant codes \\ from coding \\ history\end{tabular} & \begin{tabular}[c]{@{}l@{}}$\bullet$ \textbf{system role}: You are a helpful qualitative \\ analysis assistant,  aiding researchers in \\ developing codes that can be utilized in \\ subsequent stages, including discussions \\ for creating codebooks and final coding \\ processes;\\ $\bullet$ \textbf{user input}: Please identify the top three \\ codes relevant to this {[}text{]} from the \\ following [Code list];\\ 1. {[}Code{]} \\ 2. {[}Code{]}\\ ...\\ \\ Here is the example format of results: \\ 1. code content\\ 2. code content\\ 3. code content\end{tabular} & \begin{tabular}[c]{@{}l@{}}{[}Text{]}\\ "How A Business Works was an excellent book \\ to read as I began my first semester as a college \\ student. Although my goal is to major in Business, \\ I started my semester off with no idea of even the \\ basic guidelines a Business undergrad should know. \\ This book describes in detail every aspect dealing \\ with business relations, and I enjoyed reading it. \\ It felt great going to my additional business classes \\ prepared and knowledgeable on the subject they \\ were describing. Very well written, Professor \\ Haeberle! I recommend this book to anyone and \\ everyone who would like additional knowledge \\ pertaining to business matters."\\ \\ {[}Code list{]}\\ 1. Detailed introduction to business relations\\ 2. Inspiring guide to improve life\\ 3. Journey of light and love.\\ 4. Easy to read, highlight-worthy\\ 5. Well-written lesson on simplicity\\ 6. Rodriguez tells truth, Pelosi lies\\ \\ \textbf{Three relevant codes to} {[}\textbf{Text}{]} \textbf{from} {[}\textbf{Code list}{]}:\\ 1. Detailed introduction to business relations\\ 2. Easy to read, highlight-worthy.\\ 3. Well-written lesson on simplicity.\end{tabular} \\ \hline

\end{tabular}}
\end{table*}

\begin{table*}[!htbp]
\caption{The prompts utilized in \diffcoder in Phase 2 when communicating with the ChatGPT API to produce code suggestions for text.}
\label{tab:system:prompt2}
\scalebox{0.8}{\begin{tabular}{|l|l|l|l|}
\hline
Phases & Features & Prompt Template & Example \\ \hline
Phase 2 & \begin{tabular}[c]{@{}l@{}}Make code \\ decisions\end{tabular} & \begin{tabular}[c]{@{}l@{}}$\bullet$ \textbf{system role}: You are a helpful qualitative \\ analysis assistant,  aiding researchers in \\ developing final codes that can be utilized \\ in subsequent stages, including final coding \\ processes;\\ $\bullet$ \textbf{user input}: Please create three concise, \\ non-repetitive, and general six-word code \\ combinations for the {[}text{]} using [Code1] \\ and [Code2]:\\ 1. {[}Code{]} \\ 2. {[}Code{]}\\ ...\\ \\ Requirements: \\ 1. 6 words or fewer; \\ 2. No duplicate words;\\ 3. Be general; \\ 4. Three distinct versions\\ \\ Here is the format of results: \\ Version1: code content\\ Version2: code content \\ Version3: code content \end{tabular} & \begin{tabular}[c]{@{}l@{}}{[}Text{]}\\ "How A Business Works was an excellent book \\ to read as I began my first semester as a college \\ student. Although my goal is to major in Business, \\ I started my semester off with no idea of even the \\ basic guidelines a Business undergrad should know. \\ This book describes in detail every aspect dealing \\ with business relations, and I enjoyed reading it. \\ It felt great going to my additional business classes \\ prepared and knowledgeable on the subject they \\ were describing. Very well written, Professor \\ Haeberle! I recommend this book to anyone and \\ everyone who would like additional knowledge \\ pertaining to business matters."\\ \\ {[}Code1{]}: \\ Detailed introduction to business relations.\\ {[}Code2{]}: \\ Comprehensive guide to business basics\\ \\ \textbf{Three suggestions for final codes:}\\ Version1: In-depth overview of business fundamentals\\ Version2: Thorough guide to business relationships\\ Version3: Comprehensive resource on business essentials\end{tabular} \\ \hline
\end{tabular}}
\end{table*}

\begin{table*}[!htbp]
\caption{The prompts utilized in \diffcoder in Phase 3 when communicating with the ChatGPT API to produce code group suggestions for final code decisions.}
\label{tab:system:prompt3}
\scalebox{0.8}{\begin{tabular}{|l|l|l|l|}
\hline
Phases & Features & Prompt Template & Example \\ \hline
Phase 3 & \begin{tabular}[c]{@{}l@{}}Generate \\ code groups\end{tabular} & \begin{tabular}[c]{@{}l@{}}$\bullet$ \textbf{system role}: You are a helpful qualitative \\ analysis assistant, aiding researchers in \\ generating final code groups/main themes\\ based on the [Code list] provided, in order \\ to give an overview of the main content \\ of the coding. \\ \\ $\bullet$ \textbf{user input}: Organize the following \\ {[}Code list{]} into 3 thematic groups without \\ altering the original codes, and name \\ each group:\\ 1. {[}Code{]} \\ 2. {[}Code{]}\\ ...\\ \\ Here is the format of the results:\\ Group1: {[}theme{]}\\ 1.{[}code{]}\\ 2.{[}code{]}\\ 3.{[}code{]}\end{tabular} & \begin{tabular}[c]{@{}l@{}}[Code list]:\\ 1. Simplified business knowledge\\ 2. Unconventional, but valuable business insights.\\ 3. Effective lessons on simplicity\\ 4. Innovative leadership through Jugaad.\\ 5. Cautionary book on costly Google campaigns.\\ 6. Timeless love principles improve business.\\ 7. Politicians deceive for political gain.\\ 8. A high school must-read for financial literacy.\\ 9. Entertaining and educational graphic novel.\\ \\ \textbf{Three code groups for the above} {[}\textbf{Code list}{]}: \\ Group1: Simplified business knowledge\\ 1. Simplified business knowledge\\ 2. Effective lessons on simplicity\\ 3. Cautionary book on costly Google campaigns.\\ \\ Group2: Inspiring and practical personal development book\\ 1. Timeless love principles improve business.\\ 2. A high school must-read for financial literacy.\\ 3. Entertaining and educational graphic novel.\\ \\ Group3: Unconventional, but valuable business insights\\ 1. Innovative leadership through Jugaad.\\ 2. Unconventional, but valuable business insights.\\ 3. Politicians deceive for political gain.\end{tabular} \\ \hline
\end{tabular}}
\end{table*}

\begin{table*}[!htbp]
\caption{Demographics of Participants in User Evaluation. Note: QA expertise is not solely determined by the number of QA experiences, but also by the level of QA knowledge. This is why some participants with 1-3 times of prior experience may still regard themselves as having intermediate expertise.}
\label{tab:appendix:participants}
\scalebox{0.8}{\begin{tabular}{|ll|l|l|l|l|l|l|l|}
\hline
\multicolumn{2}{|l|}{\textbf{Pairs}} & \textbf{English} & \textbf{Job} & \textbf{Education} & \textbf{\begin{tabular}[c]{@{}l@{}}\added{Related experience}\end{tabular}} & \textbf{\begin{tabular}[c]{@{}l@{}}Self-reported \\ QA expertise\end{tabular}} & \textbf{QA times} & \textbf{\begin{tabular}[c]{@{}l@{}}Software \\ for QA\end{tabular}} \\ \hline
\multicolumn{1}{|l|}{\multirow{2}{*}{Pair 1}} & P1 & Proficient & Student & Master & \begin{tabular}[c]{@{}l@{}}Basic understanding \\ of qualitative research \\ method\end{tabular} & No Experience & None & None \\ \cline{2-9} 
\multicolumn{1}{|l|}{} & P2 & First language & \begin{tabular}[c]{@{}l@{}}Automation QA \\ Engineer\end{tabular} & Undergraduate & Automation & No Experience & None & None \\ \hline
\multicolumn{1}{|l|}{\multirow{2}{*}{Pair 2}} & P3 & First language & Phd Student & PhD and above & HCI & Expert & 7 times above & \begin{tabular}[c]{@{}l@{}}Atlas.ti \\ Desktop\end{tabular} \\ \cline{2-9} 
\multicolumn{1}{|l|}{} & P4 & First language & Undergraduate & Undergraduate & \begin{tabular}[c]{@{}l@{}}Business analytics \\ with Python and R\end{tabular} & No Experience & None & None \\ \hline
\multicolumn{1}{|l|}{\multirow{2}{*}{Pair 3}} & P5 & Proficient & Student & Undergraduate & Coding with Python & Beginner & 1-3 times & None \\ \cline{2-9} 
\multicolumn{1}{|l|}{} & P6 & First language & \begin{tabular}[c]{@{}l@{}}Research \\ Assistant\end{tabular} & Master & Asian studies & Expert & \begin{tabular}[c]{@{}l@{}}7 times above \\ (mainly interview \\ data)\end{tabular} & \begin{tabular}[c]{@{}l@{}}Word, \\ Excel, \\ Dedoose\end{tabular} \\ \hline
\multicolumn{1}{|l|}{\multirow{2}{*}{Pair 4}} & P7 & First language & Data Analyst & Undergraduate & Data Visualisation & No Experience & None & None \\ \cline{2-9} 
\multicolumn{1}{|l|}{} & P8 & First language & Student & Undergraduate & \begin{tabular}[c]{@{}l@{}}R, HTML/CSS, \\ Market research\end{tabular} & Beginner & 1-3 times & R \\ \hline
\multicolumn{1}{|l|}{\multirow{2}{*}{Pair 5}} & P9 & First language & Research assistant & Undergraduate & \begin{tabular}[c]{@{}l@{}}Learning science, \\ Grounded theory\end{tabular} & Intermediate & 4-6 times & NVivo \\ \cline{2-9} 
\multicolumn{1}{|l|}{} & P10 & First language & \begin{tabular}[c]{@{}l@{}}Data science \\ intern\end{tabular} & Undergraduate & \begin{tabular}[c]{@{}l@{}}Computer Vision, \\ Python\end{tabular} & No Experience & None & None \\ \hline
\multicolumn{1}{|l|}{\multirow{2}{*}{Pair 6}} & P11 & First language & \begin{tabular}[c]{@{}l@{}}Behavioral \\ Scientist\end{tabular} & Undergraduate & \begin{tabular}[c]{@{}l@{}}Psychology, \\ Behavioral Science, \\ Thematic analysis\end{tabular} & Intermediate & 1-3 times & Word \\ \cline{2-9} 
\multicolumn{1}{|l|}{} & P12 & First language & Student & Undergraduate & \begin{tabular}[c]{@{}l@{}}Accounting \& \\ Python, SQL\end{tabular} & No experience & None & None \\ \hline
\multicolumn{1}{|l|}{\multirow{2}{*}{Pair 7}} & P13 & First language & \begin{tabular}[c]{@{}l@{}}Research \\ Assistant\end{tabular} & Undergraduate & \begin{tabular}[c]{@{}l@{}}SPSS, Python, basic \\ qualitative analysis \\ understanding, \\ topic modeling\end{tabular} & Beginner & 1-3 times & None \\ \cline{2-9} 
\multicolumn{1}{|l|}{} & P14 & First language & \begin{tabular}[c]{@{}l@{}}Research \\ Assistant\end{tabular} & Undergraduate & \begin{tabular}[c]{@{}l@{}}Have research \\ experience using \\ QA for interview \\ transcription\end{tabular} & Intermediate & 7 times above & \begin{tabular}[c]{@{}l@{}}NVivo, \\ Excel\end{tabular} \\ \hline
\multicolumn{1}{|l|}{\multirow{2}{*}{Pair 8}} & P15 & First language & Researcher & Master & \begin{tabular}[c]{@{}l@{}}Thematic analysis \\ for interview, \\ literature review\end{tabular} & Beginner & 1-3 times & fQCA \\ \cline{2-9} 
\multicolumn{1}{|l|}{} & P16 & First language & Student & Master & Social science & No experience & None & None \\ \hline
\end{tabular}}
\end{table*}

\begin{table*}[!htbp]
\caption{Observation notes for Pair1-Pair4. The language has been revised for readability.}
\label{tab:observation_notes_1}
\scalebox{0.7}{\begin{tabular}{|l|l|l|}
\hline
 & \textbf{Atlas.ti Web} & \textbf{CollabCoder} \\ \hline
P1xP2 & \begin{tabular}[c]{@{}l@{}}Even individuals familiar with Google Docs/Excel might find it \\ challenging to adapt to Atlas.ti, P1's learning pace was even slower \\ than P2's. P1's coding was more detailed and extensive than P2's, \\ making his codes longer and more content-rich. His lack of experience \\ in ML further hampered his speed, causing a significant delay in \\ the coding process. In a 30-minute span, he managed only 5 codes \\ compared to another coder who completed all 20. Due to time \\ constraints, we had to stop the process.\\      \\ Over time, the involvement of the other coder diminished. \\ Only one coder,  more adept with the platform, primarily handled \\ the coding process. The other coder, like P1, shifted to a supportive \\ role, offering input on the final report and the categorization phase.\end{tabular} & \begin{tabular}[c]{@{}l@{}}This time around, P1 found it easier to start coding. Both he \\ and the other coder seldom used the "Similar codes" function. \\ Additionally, they rarely used the "certainty" button, indicating \\ a potential issue of over-reliance on certain features or methods.\end{tabular} \\ \hline
P3xP4 & \begin{tabular}[c]{@{}l@{}}Even the expert coder (P3) faced challenges learning the software \\ and initially felt lost navigating its interface. Additionally, she \\ found it difficult to identify the origin of selected text when only \\ a portion is chosen from the original unit.\\      \\ In both coding sessions, the lead coder shares her screen and \\ invites the other coder to offer suggestions for combining codes \\ and arriving at the final code. Due to the limitations of the software, \\ they are obliged to manually search for similar codes, relying on \\ visual inspection to group them together.\end{tabular} & \begin{tabular}[c]{@{}l@{}}P3 is a conscientious coder who is concerned about potentially \\ slowing down the overall pace of the study. To address this, she \\ intermittently checks the progress of others to adjust her own \\ workflow. She finds this feature to be "quite helpful."\\      \\ When it comes to coding, if the codes are identical, they typically \\ don't consult definitions. While ChatGPT serves as a reference point \\ for decision-making, it is not strictly followed. If there's a difference \\ in understanding, they will refer to ChatGPT for final decision support.\\      \\ When P4 sets the certainty level to 2, it signifies "I'm not sure what \\ this person is talking about." The lead coder is conscious of not overly\\ relying on her own codes, as she doesn't want to appear too dominant \\ within the team. By using third-party codes, she aims to maintain a \\ more balanced influence. P4 also mentioned that he sometimes assigns \\ low priority to definitions because he has only a few to refer to.\\      \\ During the decision-making process, direct selections from ChatGPT \\ suggestions have become less frequent. Instead, the team tends to use \\ ChatGPT is more of a point of reference. This seems to indicate that they \\ are becoming more cautious in their approach.\end{tabular} \\ \hline
P5xP6 & \begin{tabular}[c]{@{}l@{}}Beginners have the option of referring to others' codes as a \\ starting point for their own coding endeavors. P6, for instance, \\ prefers to check the code history. This approach can provide \\ valuable insights and context, helping new coders understand the \\ coding process better and potentially speeding up their learning \\ curve. The team takes advantage of the auto-completion function, \\ typing in just a few words and then clicking the check button to \\ select existing codes instead of creating new ones. When P5 is \\ coding, he initially refers to other people's codes before adding \\ his own.\\      \\ The codes generated by both coders tend to be rather general. \\ They often refer to each other's work, with the beginner usually \\ following the coding scheme established by the more experienced \\ coder.\end{tabular} & \begin{tabular}[c]{@{}l@{}}Russell is not familiar with the new coding method and initially \\ selected the entire text as "keywords support", thinking that only \\ the selected portion would be coded. This suggests that users may \\ need some training to effectively use the coding system.\end{tabular} \\ \hline
P7xP8 & \begin{tabular}[c]{@{}l@{}}To speed up the coding process,  only one coder takes on the \\ responsibility of combining and grouping the codes into thematic \\ clusters.\end{tabular} & \begin{tabular}[c]{@{}l@{}}P7 and P8 both tend to use ChatGPT sparingly, favoring the creation of \\ their own codes. P7 mentions that the long latency for ChatGPT's suggestions \\ is a factor; if the results aren't quick, he opts to input his own codes. P8 \\ notes that she often has to edit ChatGPT's suggestions, deleting some \\ words to better tailor them to her needs.\\      \\ However, they are more likely to choose suggestions from ChatGPT if they \\ want to expedite the process. Even if they don't ultimately select a ChatGPT \\ suggestion, they still refer to these codes as a reference point. This approach \\ aligns with the sentiment that AI can't be trusted for every task; it serves \\ as a tool rather than a definitive authority.\\      \\ If there's any uncertainty about why a code is part of a specific group, or if \\ the meaning of a code within a group is unclear, they will refer back to the \\ original text during the code grouping phase for clarification.\\ \\ By highlighting keywords and listing them, the coders are able to work \\ asynchronously instead of in real-time. This approach allows each coder to \\ leave markers of their understanding, facilitating a smoother integration of \\ perspectives without the need for immediate discussion.\end{tabular} \\ \hline
\end{tabular}}
\end{table*}

\begin{table*}[!htbp]
\caption{Observation notes for Pair5-Pair8. The language has been revised for readability.}
\label{tab:observation_notes_2}
\scalebox{0.7}{\begin{tabular}{|l|l|l|}
\hline
 & \textbf{Atlas.ti Web} & \textbf{CollabCoder} \\ \hline
P9xP10 & Normal collaboration process, no specific notes & \begin{tabular}[c]{@{}l@{}}The coding process involves multiple steps: initially reading the \\ data, requesting suggestions, reviewing those suggestions, returning \\ to the raw data for another check, and then choosing or editing the code. \\ After this, keywords are added and the level of certainty is labeled.\\      \\ When it comes to merging codes, the team starts with a preliminary \\ idea for a final decision, and then consults ChatGPT to generate \\ a final, merged code.\end{tabular} \\ \hline
P11xP12 & \begin{tabular}[c]{@{}l@{}}P12 adopts a strategy of starting his coding from the last data \\ point and working his way to the top, in an effort to minimize \\ overlap and influence from P11. The pace at which each coder \\ works varies significantly: one coder is much faster than the \\ other and, consequently, contributes more to the overall workload.\\      \\ In time-sensitive situations, the quicker coder naturally takes on \\ more responsibilities than the slower coder. This dynamic could \\ have implications for the diversity and depth of coding, as the \\ faster coder's perspectives may disproportionately influence the \\ final output.\end{tabular} & \begin{tabular}[c]{@{}l@{}}A less-than-ideal scenario for discussion. The team may overly \\ rely on ChatGPT's generated decisions due to time constraints. \\ In these cases, substantive discussion is replaced with shortcuts \\ like simply choosing "the first one" or "the second one" from \\ ChatGPT's suggestions. This is a notable drawback for the research, \\ as it sidesteps deeper analysis and thought, leading to concerns \\ about over-reliance on automated suggestions. The dynamic \\ often results in the more experienced coder taking a dominant \\ role in the process, which could impact the diversity of perspectives \\ in the coding.\\      \\ AI does offer the advantage of allowing users to work with \\ longer text segments compared to manual coding, which often \\ focuses on keywords or smaller data units. However, this advantage \\ should not come at the expense of thoughtful discussion and careful \\ consideration in the coding process.\end{tabular} \\ \hline
P13xP14 & Normal collaboration process, no specific notes & \begin{tabular}[c]{@{}l@{}}The overall coding process appears to be smooth. Both coders \\ generally agree and neither is overly dominant; they respect each \\ other's opinions. Because of this harmony, they utilize ChatGPT \\ to generate the final code expressions.\\      \\ It seems that AI plays a significant role in the code grouping process, \\ directing the way codes are organized. When the coders are working \\ independently, they tend to group codes based on sentiment analysis. \\ However, under AI's guidance, their focus shifts to content analysis. \\ This suggests that while AI can be a helpful tool, its influence can also \\ steer the analytical direction, which may or may not align with the \\ coders' initial approach or intentions.\end{tabular} \\ \hline
P15xP16 & \begin{tabular}[c]{@{}l@{}}Due to time constraints, discussions between the coders are \\ notably brief and to the point. There isn't much room for \\ extended dialogue or deeper analysis, which could have \\ implications for the thoroughness and quality of the coding \\ process.\end{tabular} & \begin{tabular}[c]{@{}l@{}}Participants generally start by reading the original text, then request \\ suggestions from ChatGPT before proceeding to code the data.\\      \\ P16 follows this sequence, reading the text first and then consulting \\ ChatGPT's suggestions. P15, on the other hand, sometimes deletes \\ her initial code entry to generate a different version. Due to time \\ constraints, she doesn't delve deeply into the text and may even skip \\ over some sections. To save time, she might initiate ChatGPT's code \\ suggestion process for another text segment while working on the \\ current one.\\      \\ Both P15 and P16 demonstrate mutual respect when their codes \\ closely align (with a similarity score greater than 0.9). They don't \\ particularly mind whose code is used for the final decision. For \\ instance, they may choose one of P15's codes for one text segment \\ and switch to another code from P16 for a different segment.\\      \\ If their coding doesn't match despite having similar evidence, they \\ discuss the reasons behind their code choices. The coder with the \\ more explainable rationale usually wins out, with the other coder \\ simply saying, "Use yours." If they can't reach a decision, they turn \\ to ChatGPT for a suggested code.\\      \\ When neither coder feels confident in their understanding of the \\ raw data,  they'll admit their uncertainty, often stating, "I have no \\ idea about this", before potentially seeking further guidance.\end{tabular} \\ \hline
\end{tabular}}
\end{table*}

\end{document}